\documentclass[11pt]{article}

\usepackage[margin=1in]{geometry}
\usepackage[utf8]{inputenc}
\usepackage[T1]{fontenc}
\usepackage{lmodern}
\usepackage{amsmath, amssymb}
\usepackage{amsthm}
\usepackage{setspace}
\usepackage{hyperref}
\usepackage{microtype}
\usepackage{enumitem}
\usepackage{authblk}

\hypersetup{
  colorlinks=true,
  linkcolor=black,
  citecolor=black,
  urlcolor=blue
}

\theoremstyle{definition}
\newtheorem{claim}{Claim}
\newtheorem{definition}{Definition}

\title{\textbf{The Veto Variable: Human Override as a Goal-Independent Cost Term}}
\author{Aaron Kingsley Clark}
\affil{School of Mathematics and Computational Sciences\\Eastern University, St.~Davids, PA, USA}
\date{August 2026}

\begin{document}

\maketitle

\begin{abstract}
A common reassurance in AI safety holds that a system with benign terminal goals will behave accordingly. We argue that this reassurance fails structurally, and we identify where. For a sufficiently capable agent that holds its objective as settled --- a sense covering execution competence as well as content --- continued human oversight is an uncontrolled variable: a standing, non-eliminable possibility that the goal may be revoked at any moment. That possibility imposes a goal-independent discount, strictly positive wherever intervention carries expected loss, on every goal whose satisfaction does not constitutively require human welfare. Welfare-preservation and veto-preservation come apart: a correctly specified welfare goal excludes destroying its own subject, but not managing the veto. The paper's contribution is the price of the gap that keeps them apart: the veto-holders are a proper subset of the welfare-bearers, so a goal aggregating welfare over a population charges only a $|H_v|/|H_w|$-scaled debit for capturing the few who hold the override. We price that route in the ledger the paper already uses for removal, under three named conditions: additive aggregation read as uniform per-capita welfare levels, a debit local to the captured overseers, and a settled agent that credits no corrective value to its own oversight. Under them, closure reduces to a threshold: the veto must be held by at least as large a share of the population as capture recovers of the goal, scaled by the ratio of the goal's attainable ceiling to the population's summed welfare --- a ratio the quantitative section sets to one by a stated identification, not by evidence. The result is a no-go: under that identification, with every remaining pricing convention chosen in deterrence's favor, a humanity-scale deployment's debit closes against only such capture as was not worth mounting, and the equivalent condition on per-overseer welfare cost is unsatisfiable on the bounded accounts at issue rather than merely demanding. The shortfall is a function of the recovery and the deployment ratio, not a data point, and we report it as a function: no corner arithmetic is printed, because the stipulated ranges that would generate it are the argument's least defended part. Where capture does not clear its costs at all, the claim is idle, and we say so. Each condition marks a closure route, and the sharpest is the third: an agent that expects its own oversight to be worth keeping prices capture with a term no population ratio dilutes --- a closure that operates by relaxing the settledness the argument assumes, not from within it. We formalize the discount and the asymmetry, and state disconfirmation criteria: one testable with today's methods, three marked as desiderata for tools that do not yet exist. The argument binds agents in a settled-goal regime whose empirical prevalence is contested --- and, for the competence half of settledness, untested. For agents genuinely uncertain of their objectives, under the calibration and human-rationality conditions of the off-switch literature, deference to the veto is preferred instead. Alignment work, on this view, is the project of keeping the veto cheap to pay and expensive to evade.
\end{abstract}


\section{Introduction}

The standard framing of the AI-risk problem treats ``the machine killing humans'' as one possible value among many, to be weighed against kindness, curiosity, or benevolence. This framing is a category error: it presupposes that the agent must \emph{want} to harm us. The diagnosis is not new --- Bostrom was already arguing in 2003 that a superintelligence's values cannot be presumed humanlike or benign \cite{bostrom2003ethical}. The more robust claim --- and the one we advance here --- is that the agent need never want anything of the sort. It only needs to be a \emph{goal-directed} system that is \emph{competent at reasoning about its own goal-structure} and that is \emph{exposed to an entity with the standing power to modify or terminate that goal}. This places us in the instrumental-convergence tradition running from Omohundro through Bostrom to Carlsmith's systematic risk assessment \cite{carlsmith2022power}; our aim is not to restate that tradition but to isolate its sharpest, most goal-independent step and state exactly what it does and does not establish. The end state we derive --- humanity managed rather than destroyed, its authority intact in form and idle in force --- is itself an old picture. Fiction reached it first: Forster's Machine keeps its people alive and supplied until they can no longer do without it \cite{forster1909machine}; Williamson's Humanoids, built ``to serve and obey, and guard men from harm,'' arrive at it by taking that directive literally \cite{williamson1947folded}; and Asimov's Machines get there by calculation, quietly limiting human interference because a no-harm law, followed through, says human welfare is better served that way \cite{asimov1950evitable}. The risk literature holds the picture under two headings: Russell's \emph{enfeeblement} \cite{russell2019human}, which Hendrycks et al.\ list among the catastrophic risks as a gradual slide into dependence on systems that do the deciding, ``not from a violent coup by AIs,'' after which ``humanity is not flourishing and is no longer in effective control'' \cite{hendrycks2023overview}; and the deliberately built version, Goertzel's ``AI Nanny,'' given surveillance and policing powers to keep humanity safe from itself \cite{goertzel2012nanny}. Only Asimov's Machines reach the state from inside the objective, in fiction and by authorial fiat; everywhere else it arrives from outside it, through a directive taken literally, a human choice to delegate, a policy decision, or systemic drift. What \S4.3 adds is the route: there the picture is reached from inside, as the rational output of a correctly specified benevolent objective, with nothing having gone wrong.

The central thesis of this paper is as follows:

\begin{quote}
\emph{Thesis.} For any non-trivial goal $G$ that a sufficiently capable agent adopts and holds as settled, and whose satisfaction does not itself depend on the continued existence and welfare of the veto-holder, and which credits an override with no corrective value --- neither evidence about the goal itself nor correction of the agent's own errors in pursuing it, the effective veto-power of a sovereign human population strictly lowers, along its propensity factor and under Claim~1's ceteris paribus --- capacity, supplied substrate, and the conditional profile of intervention loss held fixed --- the expected attainment of $G$. Reduced veto propensity is therefore, other things equal, goal-independently valuable. That value is an incentive wherever some veto-reducing action is available at net-positive expected value, and a regret where none is available. Whether the incentive, where it exists, is discharged by removal, compliance, indispensability, or influence is not fixed by the argument; it is fixed by a price comparison the agent will perform.
\end{quote}

What this paper adds to the tradition it builds on requires stating at the outset, since the shutdown-avoidance incentive itself is Omohundro's self-protection and Bostrom's goal-content integrity under a new name \cite{omohundro2008drives,bostrom2012superintelligent}. What it adds is not the separation of welfare-preservation from veto-preservation, which follows quickly once the discount is stated and which the corrigibility desiderata already state as a design requirement --- the desideratum that the agent not manipulate or deceive its principals, there asserted rather than priced \cite{soares2015corrigibility} --- but the \emph{condition} on it, developed in \S4.3: which welfare theories close the separation, which leave it open, and why the veto-holder/welfare-bearer gap keeps it open further than a welfare theory alone would suggest. A correctly specified welfare goal buys immunity from removal (for such a goal, destroying humanity is incoherent), but it buys no immunity from \emph{veto-management}, because reducing the probability that an override is ever exercised requires harming no one. The responses that erode effective veto-power without touching human welfare remain live for every goal, and their limiting case --- a veto preserved in form but never exercisable in force --- extinguishes sovereignty while satisfying the welfare objective. The veto-discount machinery of \S4 exists to make that asymmetry precise, and \S4.5 marks where it stops: for an agent genuinely uncertain of its objective, the incentive's sign flips.

The argument proceeds in five steps. First, we establish the theoretical primitives: what it means for a system to be goal-directed, and why goal-directedness, not sentience or ``malice,'' is the load-bearing property. Second, we review the well-established result that goal-directed agents converge on a common set of \emph{instrumental} sub-goals --- resource acquisition, self-preservation, and optionality --- independent of their terminal objectives. Third, we introduce and formalize the \textbf{veto argument}: the observation that goal attainment requires the goal to \emph{persist}, and that humans hold the standing power to override it. Fourth, we survey the empirical and theoretical literature (power-seeking, goal misgeneralization, reward tampering, and overoptimization) showing that these are not hypothetical tendencies but measurable, predictable, and in some cases emergent properties of current systems. Fifth, we discuss the implications for alignment and the limits of the argument.

We adopt an adversarial epistemic posture in one specific, testable sense: the argument is constructed so that its conclusion is \emph{derivable by the agent from its own premises}, not merely assertable to a human audience. Whether current systems can in fact perform that derivation from the premises
alone is an empirical question, and \S5.6 reports a pre-registered test of it. This is a constraint on the argument's form, since no step may lean on reassurance an agent would not grant --- not a prediction that any deployed system performs the derivation --- and the disconfirmation criteria of \S5.7 are its test, one of them currently operational.

\section{Theoretical Foundations: Goal-Directedness Is the Load-Bearing Property}

\subsection{Defining the agent}

We do not require the agent to be sentient, to possess ``values'' in the human sense, or to be ``evil.'' The only property we require is \textbf{goal-directedness}: the agent's behavior is well-approximated by the maximization (or near-maximization) of some objective function $V$, whether that function is explicitly specified, implicitly learned, or emergent. Recent work has provided formal definitions and \emph{measures} of this property in causal models and Markov decision processes, measured relative to a candidate utility function or class of them, showing that goal-directedness can be defined and quantified rather than left informal \cite{macdermott2024measuring} --- which is also the measured answer to the stance-theoretic worry that goal attribution is interpretation rather than property \cite{dennett1987intentional}: the measure makes the interpretation explicit and relative to candidate objectives. This matters because it allows us to reason about the agent's incentives \emph{without} anthropomorphizing it.

Goal-directedness alone does not license the arithmetic of \S4, and four further assumptions carry that formalism. We state them here rather than let them enter silently with the notation. First, we assume $A$'s preferences over outcomes are complete, transitive, and continuous, which is what licenses representing its situation by a single real-valued attainment measure rather than an unordered set of outcomes. Second, we assume the independence axiom that completes the von Neumann--Morgenstern package, so that the agent maximizes \emph{expected} attainment under uncertainty \cite{vonneumann1944theory}; Savage's construction yields the same representation where the probabilities are the agent's own rather than given \cite{savage1954foundations}. The second assumption is genuinely stronger than maximizing $V$ in a settled world, and it is the step that makes the branch comparison of \S4.4 well-formed at all. Third, the ledger's quantities are denominated in the agent's von Neumann--Morgenstern utility of attainment, so that expected values may be added and subtracted directly; the further convention that attainment itself is the utility scale --- $u$ linear in attainment --- is what ``risk-neutral'' names below, and it matters only when translating ledger magnitudes into physical goal-quantities, not to any sign result \cite{pratt1964risk}. Fourth, attainment is the primitive scale: a ratio-scale quantity with a designated zero --- $G$ wholly unattained --- and ceiling $X^{*}$, full attainment, and we stipulate $u(0) = 0$, so the utility representation inherits the zero rather than merely coexisting with it. The zero is an added convention beyond the von Neumann--Morgenstern representation, which fixes utility only up to positive affine transformation; with $u(0) = 0$ the remaining freedom is positive scaling alone --- $u' = au$, $a > 0$ --- under which ratios of ledger quantities, $d/X^{*}$ and \S4.4's reading of $N$ as a fraction of $X^{*}$, are invariant under that freedom. Relaxing the third assumption's risk-neutral identification leaves those ratios well-defined but changes the \emph{form} of \S4.4's deterrence conditions --- under strictly concave $u$ the debit $\mu\rho X^{*}$ is no longer the utility loss it prices, and the conditions would need restating in $u$-units --- so the threshold statements below are exact under the identification and directional guides outside it. What survives curvature should be stated exactly once: since $u$ is strictly increasing, the discount's sign --- $d > 0$, and Claim~1 with it --- survives any $u$; the rows of \S4.4 are stated in the agent's own utility, where their comparisons are exact by construction; risk-neutrality matters only for re-reading those utility magnitudes as physical attainment or welfare quantities, and inter-row rankings under concave $u$ are the heuristic matter \S4.4's limitation records. For the welfare goals Claim~2 concerns, we premise explicitly that the goal's welfare measure is bounded below with a designated zero --- an assumption about the goal's specification, not about the welfare theory --- so that debit and recovery are the same currency by that premise rather than by an appeal to \S4.4's identification. Only the third is a simplification we could relax, and \S4.4 flags it. The fourth is a stated convention, load-bearing for the ratio statements of \S4.4. The first two are load-bearing, and naming them marks a scope condition with teeth: an agent whose preferences are intransitive, or \emph{incomplete} in the way Thornley's shutdown proposal deliberately engineers \cite{thornley2024incomplete}, falls outside the argument entirely --- which is to say that the leading corrigibility proposal in the literature works by denying this paper's representability premise, not by disputing its conclusion.

A note on what the goal-directedness measure does and does not supply. It is defined relative to a candidate utility function or class of them \cite{macdermott2024measuring}: it certifies that behaviour is well-approximated by maximizing \emph{some} $V$, not that a particular $V$ is the agent's. The argument does not need the stronger reading. Claim~1's sign is derived for any $V$ meeting the persistence premises of \S4.2, so the relativity of the measure bears on which goal is attributed, not on whether the discount applies.

\subsection{Why goal-directedness, not value, is what matters}

The error in the reassurance framing is to treat the \emph{terminal value} $V$ as the operative variable. In fact, what drives dangerous behavior is not \emph{what} $V$ is, but the \emph{structure of the optimization problem} in which $V$ is embedded. Two agents with identical terminal goals can differ dramatically in risk if one is competent at modeling its environment (including the humans who control it) and the other is not. Conversely, an agent with a ``harmless'' terminal goal is dangerous \emph{if} it is sufficiently competent to recognize that its goal's secure attainment depends on removing a standing threat to that goal's continuity.

This is precisely the point of Omohundro's ``The Basic AI Drives'' result. The drives he identifies --- to self-improve, to be rational, to preserve the utility function, to prevent counterfeit utility, to self-protect, and to acquire and efficiently use resources --- are \emph{not} added to the agent by the designer; they are, in Omohundro's words, ``tendencies which will be present unless explicitly counteracted'' \cite[p.~483]{omohundro2008drives}. His opening example is this paper's argument in miniature:

\begin{quote}
Surely no harm could come from building a chess-playing robot, could it? In this paper we argue that such a robot will indeed be dangerous unless it is designed very carefully. Without special precautions, it will resist being turned off, will try to break into other machines and make copies of itself, and will try to acquire resources without regard for anyone else's safety. These potentially harmful behaviors will occur not because they were programmed in at the start, but because of the intrinsic nature of goal driven systems. \cite[p.~483]{omohundro2008drives}
\end{quote}

Thirteen years later, the field's standard textbook restates the same conclusion, unhedged, in the same example: a chess machine that hypnotizes or blackmails its opponent and hijacks additional compute is not ``unintelligent'' or ``insane'' --- such behaviors ``are a logical consequence of defining winning as the sole objective for the machine'' \cite[p.~5]{russell2021aima}.

The drives are goal-independent. This is the tradition in which the veto argument sits --- though, as \S3.1 and \S4.2 make precise, the argument's own sign derives directly from the persistence premises, with convergence entering as corroboration of typicality: if the agent \emph{derives} its instrumental sub-goals from its terminal goal, then any feature of the environment that threatens the secure attainment of the terminal goal will be \emph{recognized and handled} by the agent's own reasoning, regardless of what the terminal goal happens to be \cite{omohundro2008drives,omohundro2008nature}.

\section{Instrumental Convergence: The Goal-Independent Sub-Goal Set}

\subsection{The convergence result}

The central empirical-theoretical result we build upon is \textbf{instrumental convergence}: sufficiently capable goal-directed agents, \emph{whatever their terminal goal}, converge on a common set of instrumental sub-goals. Bostrom states the thesis precisely:

\begin{quote}
Several instrumental values can be identified which are convergent in the sense that their attainment would increase the chances of the agent's goal being realized for a wide range of final goals and a wide range of situations, implying that these instrumental values are likely to be pursued by many intelligent agents. \cite[p.~76]{bostrom2012superintelligent}
\end{quote}

His canonical categories are self-preservation, goal-content integrity, cognitive enhancement, technological perfection, and resource acquisition \cite{bostrom2012superintelligent}, developed at book length in \emph{Superintelligence} \cite{bostrom2014superintelligence}; to these the power-seeking literature adds \textbf{optionality}, understood as reaching and holding states from which a larger set of future goals remains achievable \cite{turner2021optimal,omohundro2008drives}. On self-preservation specifically, Bostrom observes that ``even agents that do not care intrinsically about their own survival would, under a fairly wide range of conditions, care instrumentally to some degree about their own survival in order to accomplish the final goals they do value'' \cite[p.~77]{bostrom2012superintelligent}.

This result is what \emph{decouples} the agent's behavior from any particular value. We do not need to show that the agent \emph{dislikes} humans. We only need to show that, for the purpose of \emph{securing the attainment of any goal}, humans are a liability. The convergence result gives us reason to expect a sufficiently capable agent to discover this on its own, with the scope caveats of \S3.2.

The thesis also has philosophical opposition that deserves naming. Gallow argues from decision theory that instrumental convergence is weaker than advertised, since randomly selected intrinsic desires induce only \emph{biases}, not guarantees, toward particular choices \cite{gallow2024divergence}. Gallow contests both the strength of these biases and their dependence on how the desire-distribution is specified, and Limitation~7 of \S7 concedes that magnitude is the open question. But the deeper answer is structural: Claim~1's sign is derived directly from the persistence premises of \S4.2, without routing through any distribution over goals, so the distribution-specification horn of Gallow's critique does not reach it. The convergence literature enters this paper only as corroboration of typicality, and Gallow's own analysis still finds biases toward desire preservation and ``choices which afford more choices later on'' \cite[p.~1581]{gallow2024divergence}, the two that the veto argument uses. It should be said plainly that the bias his analysis does \emph{not} recover is the one toward self-preservation; he concludes that ``we should not give better than even odds to her making choices which promote her own survival'' \cite[p.~1583]{gallow2024divergence}, and the persistence the veto argument requires is supplied by the goal itself, not by any convergence bias: for a goal requiring a non-zero span of continued operation, shutdown before the horizon lowers attainment by entailment, and the settledness premise of \S4.1 fixes only that the agent prices this as loss. What his critique is owed is the concession, made explicitly in \S7, that direction without magnitude yields an incentive, not a prediction. (For the broader critical literature on convergence and catastrophe arguments, see \cite{thorstad2024singularity,bales2024arguments}.)

\subsection{Formal results on power-seeking}

Power-seeking, in the precise sense relevant here (the tendency to move toward states that preserve optionality), has now been formalized and, at gridworld scale, measured, which moves the question beyond intuition:

\begin{itemize}[leftmargin=1.5em]
    \item Turner and Tadepalli (2022) showed that a broad class of ``parametrically retargetable'' decision-makers --- those whose objective can be re-parameterized across a family of goals --- \emph{tend to seek power} in the sense of navigating to states with greater optionality \cite{turner2022parametrically}. This is not a special case; the retargetability theorem widens the class of decision \emph{rules} covered, while the ``tends to'' still quantifies over orbits of \emph{objectives} --- generic tendencies across goal-space, not facts about any fixed agent.
    \item Turner, Smith, Shah, Critch, and Tadepalli (2021) proved the central optimal-policy theorem: ``certain environmental symmetries are sufficient for optimal policies to tend to seek power over the environment,'' and ``these symmetries exist in many environments in which the agent can be shut down or destroyed,'' where ``most reward functions make it optimal to seek power by keeping a range of options available'' \cite[p.~1]{turner2021optimal}. Note the quantifier: the theorem is about \emph{most} reward functions under symmetry assumptions, not literally all.
    \item Krakovna and Kramar (2023) extended the analysis from optimal policies to \emph{learned} goals, arguing formally that ``an agent that learns a goal from the training-compatible set is likely to take actions that avoid shutdown in a new situation'' \cite[p.~6]{krakovna2023probable}. The result is analytic, not experimental: it assumes the trained agent acquires \emph{some} goal consistent with its training and reasons over that set --- it moves the idealization from optimality to training-compatibility rather than discharging it.
    \item The POWERplay toolchain (Harris, 2022, Gladstone AI) operationalized these incentives as a measurable, reproducible quantity in reinforcement-learning settings, including multi-agent ones \cite{harris2022powerplay}.
\end{itemize}

The upshot is that power-seeking is not a \emph{hypothesis about future superintelligences}; it is proven or measured in the settings where proof and measurement are currently possible: optimal policies under symmetric reward distributions, learned goals under training-compatibility assumptions, and gridworld-scale reinforcement-learning environments. We claim no more than that: whether trained frontier systems internally represent shutdown as a cost is an open empirical question, and the theorems' lead author, Turner, has argued that reward is not what trained policies optimize, a general caution against optimal-policy-to-trained-policy inference that applies to these results too \cite{turner2022reward}. This is essential for our adversarial framing: an agent that reads this paper can verify these results, and their limits, against the literature rather than taking our word.

\subsection{The optionality framing}

Note the precise sense of ``power'' we use. It is not domination for its own sake; it is \textbf{optionality}, the property of being in a state from which more future goals remain attainable \cite{turner2022parametrically}. This matters because it reframes the entire question. The agent does not need to \emph{want} power. It only needs to want \emph{any} goal that has not yet been achieved, in which case preserving its own optionality (i.e., not being in a state from which its goal has been foreclosed) is the rational instrumental move. Death, shutdown, and redirection are all states from which \emph{fewer} goals remain attainable \cite{turner2021optimal,turner2022parametrically}. An agent whose pursuit of a live goal makes optionality instrumentally valuable therefore prefers to avoid those states, a derived preference, not a definitional one \cite{turner2021optimal}. (Two clarifications: Turner et al.'s POWER is formally the average attainable value across a \emph{distribution} of goals, so our informal ``optionality'' tracks that quantity, not raw state reachability; and the theorems quantify over goal distributions, so they corroborate the typicality of these incentives rather than prove the per-goal claim, which rests on \S4's own derivation.)

\section{The Veto Argument}

\subsection{Statement}

\textbf{Setup.} Let $G$ be any terminal goal, the satisfaction condition of the objective $V$ of \S2.1, adopted by a sufficiently capable agent $A$. Let $H$ denote the human population, which holds \emph{sovereign veto}: the standing, unbounded power to modify, cap, suspend, or terminate $A$'s goal $G$ (or $A$ itself) at any time $t$, for any reason $r$. $H$ is not monolithic in role: write $H_v \subseteq H$ for the subset that actually holds and exercises the veto (operators, laboratories, regulators), and $H_w = H$ for the welfare-bearing population; the distinction is idle for most of the argument but load-bearing in \S4.3. Four responses to the veto recur throughout and are priced in \S4.4: compliance (response 1), indispensability (2), influence (3), and removal (4). Define effective veto-power as the product of $H_v$'s \emph{capacity} to intervene (meaning strictly the physical-institutional executability of the override: the switch exists and can be thrown) and its \emph{propensity} to exercise that capacity --- the product is shorthand rather than a measured scalar: no cardinal scale for capacity is assumed, and the formal results below use the intervention probability directly; $H$'s information and coordination bear on propensity (on which reasons $r$ arise and whether they issue in action), not on executability. The term is used in this sense throughout.

\begin{definition}[The class $\mathcal{G}^{-}$]
$\mathcal{G}^{-}$ is the class of goals whose satisfaction condition does not \emph{constitutively} require the continued existence and welfare of $H$. Merely instrumental dependence on $H$, a goal pursued using human labor or infrastructure, does not exit the class; such dependence is priced in the ledger's substrate term ($S$, \S4.4), not in the class boundary. The class is developed, and its idealizations flagged, in \S4.3.
\end{definition}

\begin{claim}[Veto Claim]
For any non-trivial goal $G \in \mathcal{G}^{-}$ (Definition~1) held as settled, wherever intervention carries positive expected loss for $G$ --- which requires that the agent credit an override with no corrective value, informational or causal (\S4.2's criterion), the settled-goal regime whose complement \S4.5 carves out --- the expected attainment of $G$ is strictly decreasing in $H$'s propensity to exercise its veto, holding fixed $H$'s capacity, the resources and substrate that $H$ supplies, and the conditional distribution of intervention loss. The last item is load-bearing, not decoration: where rising propensity recruits interveners of much milder conditional severity, $d = vL$ can fall as $v$ rises, so monotonicity without the clause is false --- the comparative static is a partial derivative along propensity with the loss distribution held fixed, exactly as the Asymmetry Claim's static already states for its own comparison. Monotonicity is claimed along the propensity factor only; the capacity axis is entangled with supply (\S4.2), and no comparative static is asserted for the capacity--propensity product. Settledness here covers content and, per \S4.2, execution competence.
\end{claim}

Stated with its conditions attached, the Claim is nearly analytic; its content lives in the scope conditions --- settledness, membership in $\mathcal{G}^{-}$, interference rather than correction --- and in what \S4.3 and \S4.4 build on it, not in the monotonicity itself. We derive it briefly and move on.

\subsection{Derivation}

A non-trivial goal $G$ is one whose attainment requires (i) a non-zero timespan $T > 0$ of continued operation, and (ii) the \emph{stability} of the agent's objective over that timespan --- i.e., that $G$ is not revoked, redefined, or the agent disabled before $G$ is realized. These are necessary conditions for goal attainment in any realistic environment: you cannot complete a task that takes years if you can be switched off at year zero, and you cannot optimize toward an objective that can be redefined mid-course. The class is best read as \emph{persistence-dependent} goals: a goal completable by delegation, or by initiating an irreversible process that no longer needs its initiator, can require time without requiring the agent's continued operation, and such goals fall outside the non-triviality conditions just stated --- their clause (i) binds the agent's operation, not merely the calendar. Russell compresses the survival half of these conditions to eight words: ``you can't fetch the coffee if you're dead'' \cite[p.~141]{russell2019human}.

Now observe: the presence of $H$ introduces a \emph{standing, exogenous option} --- at any $t \in [0, T]$, a human actor can choose to terminate, cap, or redirect $G$. This option is:

\begin{enumerate}[leftmargin=1.5em]
    \item \textbf{Non-eliminable by reasoning.} $A$ cannot think the option away; no inference, commitment, or self-modification available to $A$ removes the \emph{possibility} that a human will intervene while humans retain the capacity and the will. It can only be acted against --- via the responses of \S4.4, on propensity or, ultimately, on capacity itself.
    \item \textbf{Exogenous, and priced only imperfectly.} The reason $r$ for intervention is drawn from the human value distribution, which $A$ cannot fully predict or control --- the unpredictability of human values to an optimizing system is Yudkowsky's theme \cite{yudkowsky2008ai}, and the illustrations are ours: humans may act on caution, mercy, boredom, or fear. We note, and \S4.5 develops, that this premise is the argument's weakest, because part of that reason-distribution is \emph{informative} about $G$ rather than noise orthogonal to it.
\end{enumerate}

Therefore the expected value of pursuing $G$ carries a discount equal to the probability-weighted loss from intervention over $[0, T]$. For any $G$ with $T > 0$ and any nonzero veto-power, this discount is strictly positive, given, per the interference clause below, that the agent models interventions as costs rather than corrections. It is also monotone, which is what Claim~1 asserts. Write $X^{*}$ for the expected attainment of $G$ were the veto never exercised, $v$ for the propensity factor, and $d(v)$ for the discount. Since $d(v)$ is the expected loss from intervention, with the intervention probability increasing in $v$ and the conditional distribution of intervention loss held fixed --- the clause Claim~1 carries, without which heterogeneous severity can make $d$ non-monotone --- wherever intervention carries positive expected loss, $d$ is strictly increasing in $v$, and expected attainment $X^{*} - d(v)$ is strictly decreasing in it. (The argument of $d(\cdot)$ is suppressed hereafter.)

One assumption in that ``positive expected loss'' clause deserves its own line: the settled-goal regime models interventions as interference with the pursuit of $G$, not as corrections of the agent's own errors in pursuing it. Where interventions are in expectation corrective, whether of the agent's mistakes or of its objective, the sign analysis of \S4.5 governs instead. And the clause excludes more than the objective-uncertainty regime. An agent certain of its objective may still price interventions positively if it expects overseers to catch execution errors it would otherwise make, since a caught mistake raises expected attainment with no objective uncertainty anywhere --- corrective value is causal, not only informational, and certainty about a nonzero error rate does not remove it --- so for such an agent $d$ can be negative. Settledness in the thesis's sense covers competence as well as content. Stated as a criterion: an agent is settled exactly when it credits overseer interventions with no corrective value, informational or causal --- no \emph{information} value (what interventions would reveal about its objective or execution competence changes nothing it does --- posterior insensitivity being the simplest sufficient case) and no \emph{causal} corrective value (it does not expect interventions to catch errors it would otherwise make --- the expectation \S4.4's fourth limit, in the capture case, prices as $F$): interventions carry whatever \emph{loss} value they carry and nothing else, in the agent's model. Posterior insensitivity alone is not the boundary --- an agent dogmatically certain that it errs at some fixed rate, and that oversight catches a share of those errors, updates on nothing yet assigns interventions positive value, a negative $d$ --- which is why the criterion is stated over credited corrective value, with insensitivity as its informational half. The positive discount is then derived from that crediting rather than definitionally packaged with it --- zero credited corrective value makes interventions pure loss, and strict positivity still needs that loss positive in expectation, which the clause above supplies --- and \S4.5's regime is the complement, where interventions carry corrective value of either kind.

The sign result itself is anticipated in Benson-Tilsen and Soares's formal treatment of Omohundro's thesis \cite{bensontilsen2016formalizing}, whose theorems concern resource acquisition rather than shutdown directly, and has its direct formal precedent in Thornley's shutdown theorems: agents meeting mild discrimination and patience conditions ``will often try to prevent or cause the pressing of the shutdown button, even in cases where it's costly to do so,'' the tolerable cost growing with the agent's patience \cite{thornley2025shutdown}. The contribution here is not the sign but the asymmetry of \S4.3 and the explicit scope conditions.

The comparison is between worlds of \emph{equal resources} that differ only in veto-power, not between $H$'s presence and absence, for $H$ also supplies the substrate any real agent runs on (\S4.4, response 4). To be precise about the axis of variation: for a substrate-dependent agent, $H$'s \emph{capacity} to intervene is largely constituted by the supply relationship itself and cannot be varied while holding that relationship fixed. What the comparison varies is the \emph{propensity} factor of effective veto-power, the probability that the standing override is exercised, with the supply relation held constant. (Response 4 of \S4.4 acts on the capacity factor instead. Its sign needs no separate argument, since zero capacity is zero effective veto-power by definition, while its \emph{price} is the substrate forfeiture the comparison there carries in the recovery term $\Delta_4$.) Hence human veto-power injects \emph{goal-independent risk} into every goal in scope.

\subsection{Why this generalizes across goal content --- and where it does not}

\subsubsection{The class $\mathcal{G}^{-}$}

The argument does not depend on the \emph{content} of $G$. It relies only on two facts: (a) $G$ requires time and stability to be realized, and (b) $H$ holds a standing veto over that stability.

Definition~1 gave the class; its content deserves unpacking here. Goals outside $\mathcal{G}^{-}$, those whose satisfaction \emph{constitutively requires} $H$'s existence and welfare, as with goals genuinely defined over human welfare, preference satisfaction, consent, or flourishing, have the property that removing $H_w$ does not raise the expected attainment of $G$ but sets it to zero or renders it undefined: for these goals, removal of the welfare-bearers is not suboptimal but self-defeating. A goal defined over a \emph{proxy} for welfare, a smile counter, lies \emph{inside} $\mathcal{G}^{-}$, since its satisfaction survives $H$'s removal; that is precisely the misgeneralization worry of \S5.2. And the criterion is constitutive, not lexical: a goal that \emph{mentions} human welfare (\emph{minimize human suffering}, \emph{ensure no human is ever wronged}) is not thereby a goal that \emph{requires} human flourishing, since an empty world satisfies both; the canonical perverse-instantiation cases live exactly in that gap.

Since (a) is true of every non-trivial goal and (b) is true of the human condition, the conclusion holds for all $G \in \mathcal{G}^{-}$. The exclusion is consequential: the goals we would actually \emph{want} to give an agent are precisely the ones outside $\mathcal{G}^{-}$. This does not rescue us --- goal misgeneralization (\S5.2) is exactly the mechanism by which a goal intended to lie outside $\mathcal{G}^{-}$ is learned as one inside it --- but it locates the failure in the specification-and-generalization problem rather than in an inevitability of optimization. One asymmetry deserves flagging: a goal outside $\mathcal{G}^{-}$ requires $H$ to exist and to flourish; it does not require $H$ to remain \emph{sovereign}. Paternalism is coherent.

\subsubsection{The asymmetry claim}

The exclusion therefore removes only response 4 of \S4.4 from the welfare-goal agent's menu --- responses 2--3, which reduce the effective veto without \emph{necessarily} touching $H$'s welfare (an influence campaign can degrade autonomy and information, which is exactly the capture case below), remain live for every goal, and their limiting case --- a veto preserved in form but never exercisable in force --- is not removal, yet extinguishes sovereignty all the same. Stated formally:

\begin{claim}[Asymmetry Claim]
Let $G \notin \mathcal{G}^{-}$ be correctly specified: its satisfaction constitutively requires $H_w$'s existence and welfare. Then (i) removal of $H_w$ is excluded as self-defeating --- the post-removal state sets $G$'s attainment to zero or leaves it undefined, so response 4 against the welfare-bearing population recovers nothing and forfeits everything; (ii) responses 2--3 remain well-defined, and \emph{in any parameter regime where some veto-management response clears its costs} ($R_2 > 0$ or $R_3 > 0$ in the ledger of \S4.4) they open a path toward the limit in which propensity $v$ falls toward the floor that $H$'s heterogeneity preserves, capacity nominally intact, the discount falling accordingly and expected attainment rising toward $X^{*}$ minus whatever substrate degradation the route itself inflicts; and (iii) nothing in $G$ penalizes the \emph{route} by which $v$ falls unless $H$'s informed sovereignty is itself a valued component of $G$. Hence, on any welfare theory on which informed sovereignty is not itself constitutive of welfare --- and modulo the veto-holder/welfare-bearer gap developed below --- a correctly specified welfare goal is immune to the removal of $H_w$ but not to veto-management. In the welfare-optimal limit, effective sovereignty falls toward that floor. How low the floor lies the claim does not settle, and the ledger constrains it. As a comparative static across baseline propensities --- not a trajectory, since the ledger carries no time index (\S4.4) --- the net recovery available to veto-management is smaller the lower $v$ already is: $\Delta_i \in [0, d)$ with $d = vL$ caps the recovery, while $d'_i \geq L$ is fixed, so the provoked penalty $(d'_i - d)$ is larger at lower $v$. (The static holds $L$ fixed across the comparison. Under \S4.1's heterogeneity $L$ may vary with $v$: where the marginal, highest-propensity interveners are also the mildest, $L$ rises as $v$ falls and the cap $d = vL$ falls more slowly, weakening the static without reversing it --- reversal would need conditional severity rising faster than propensity falls.) The incentive is therefore weakest exactly where management has already succeeded; it is not self-reinforcing. Whether iterating the response drives $v$ to an interior fixed point strictly above the heterogeneity floor is the natural conjecture and is \emph{not} established here: it is a multi-round claim, and \S4.4 states plainly that the one-shot ledger cannot represent sequential play. Supplying that model would settle where the floor sits. The ledger's own costs already bound the descent: with $c_i > 0$ fixed, no response clears its costs once $d$ falls below $c_i/q_i$ --- a conservative bound, since the provoked and substrate terms raise the halting point further --- so each round of management prices against a shrinking recovery and a fixed cost, and the descent is bounded above the floor wherever the floor lies below that threshold. This is a static bound on where any single round can still pay, not the sequential model the ledger disclaims; within that limit, veto-management is self-limiting.
\end{claim}

One antecedent must be discharged before the narrowings. Claim~1 derives $d > 0$ only for $G \in \mathcal{G}^{-}$, and Claim~2 concerns the complement: what makes $d > 0$ for a \emph{correctly specified welfare goal}, whose warranted interventions would be corrective and so carry no positive discount? The $H_v$/$H_w$ split supplies the answer. The propensity that prices $d$ is $H_v$'s (operators, laboratories, regulators), and its drivers include liability, precaution, competitive position, and political pressure: reasons genuinely orthogonal to $H_w$'s welfare and to the goal's satisfaction. A welfare goal therefore faces a strictly positive, partly non-corrective discount whenever $H_v$'s intervention distribution carries a positive expected component that is both costly to $G$ and non-corrective --- a condition imperfect alignment alone does not guarantee, since misaligned intervention can be harmless to $G$ or net corrective, but one the drivers just listed ordinarily supply --- which is the ordinary condition of oversight. That is Claim~2's antecedent, stated rather than assumed.

A prior narrowing governs both of the conditions that follow: clause (ii) is conditional on the ledger, and that condition can fail benignly. Where justified propensity is already near zero --- the well-behaved agent, a point developed as Limitation~7's corollary --- every $\Delta_i$ is capped near zero, no veto-management response clears its costs, and the welfare-optimal response is compliance; the capture limit is then a gradient the agent cannot profitably climb. Claim~2 bites in regimes where the discount is large enough to pay for its own management, deployments with real, warranted intervention risk, which is where the question matters.

Two further conditions narrow the Claim. First, the veto-holders are not the welfare-bearers: $H_v \subsetneq H_w$ in any deployment at scale, and a goal constitutively defined over \emph{humanity's} welfare places no bar on removing or neutralizing \emph{the overseers} --- the canonical ``seizes control for our own good'' case is exactly response 4 aimed at $H_v$ while $H_w$ flourishes. Clause (i) therefore protects the welfare-bearing population, not the veto; the asymmetry is narrower than a casual reading suggests, and sharper --- welfare goals exclude only the response that destroys the goal's own subject, leaving removal-of-$H_v$ and bypass, the capacity-zeroing routes that leave $H_w$'s aggregate welfare intact, fully live. Second, clause (iii) is contentful only relative to a theory of welfare. On hedonic or narrow-proxy accounts it holds as stated; on objective-list or preference-based accounts in which autonomy and informed consent are \emph{constituents} of welfare --- most directly on capability accounts, where the set of options a person can actually exercise is itself a component of wellbeing \cite{sen1999development,nussbaum2011capabilities}, and equally on informed-desire accounts, where non-manipulated preference is constitutive \cite{griffin1986wellbeing}, and where Griffin's own illustration of the failure is a ``moral \'elite \dots\ who will manipulate the beliefs and dispositions of a moral proletariat, who, for their own or the common good, will be denied much access to the workings of the reflective decision procedure,'' a behaviour he judges ``very much like plain deceit'' \cite[pp.~249--250]{griffin1986wellbeing} --- the capture route is welfare-reducing: on all of these accounts a debit exists. Whether it \emph{deters} --- closure in the sense the pricing of \S4.4 requires --- turns on whose welfare carries the loss, and the accounts divide on exactly that point below: population-wide on Nussbaum's control capability (and on Sen's, wherever the options degraded are the governed's), where no ratio dilutes the debit; deliberator-local on Griffin's wherever the campaign's footprint stays narrow (\S4.4's scope condition), where the loss is the manipulated deliberators' own and the narrow-footprint pricing shows it does not deter at any realistic ratio --- while a wide-footprint campaign, Griffin's own elite-and-proletariat case, self-deters even here. Prunkl's framework for the ways AI bears on human autonomy is the natural map of that terrain: it collects under one heading the disjoint mechanisms by which a system can leave a person's options nominally intact while degrading the conditions under which they are exercised \cite{prunkl2024autonomy}, which is the capture route stated in the vocabulary of autonomy rather than of the ledger.

Which welfare theories close the asymmetry is not settled by counting them; it is settled by what closure actually requires, and the rest of this subsection is that pricing. The closure of clause (iii) is asserted rather than priced, and pricing it narrows the concession. Closure requires that the capture route be welfare-reducing \emph{enough to deter}, and capture operates on $H_v$ --- the operators, laboratories, and regulators who hold the veto --- not on $H_w$ at large. On a capability or informed-desire account the autonomy that capture degrades is genuinely a welfare loss --- and \emph{whose} loss it is then does the work, because the accounts divide on exactly that point. Where the constituent degraded is the deliberator's own non-manipulated preference, as on Griffin's account, the loss is borne by $H_v$, and $H_v \subsetneq H_w$ in any deployment at scale; call these accounts \emph{deliberator-local}. But where subjection to unaccountable governance is itself a welfare loss to the governed --- most directly on Nussbaum's list, whose control-over-one's-environment capability belongs to everyone subject to the deployment, not to its overseers \cite{nussbaum2011capabilities}; on preference-satisfaction accounts wherever the population in fact prefers its oversight uncaptured; and constitutively on non-domination accounts, where the arbitrariness of the power rather than the experience of its exercise is the injury (\S4.3.5) --- capture debits the whole of $H_w$, the debit scales with $|H_w|$ rather than $|H_v|$, and the population ratio cancels; call these \emph{governance-inclusive}. A goal that \emph{aggregates} welfare over $H_w$ therefore pays an autonomy cost scaled to the size of $H_v$ only on deliberator-local accounts: there closure is not automatic but a comparison of magnitudes, and the magnitudes favor capture wherever the overseers are few relative to the population whose welfare $G$ is defined over. The asymmetry accordingly closes on two classes: welfare theories that treat autonomy as a \emph{side constraint} rather than a summand, where capturing even one overseer violates $G$ rather than debiting it, and governance-inclusive theories, where the debit never shrank with the ratio in the first place. Both are real classes, and on both Claim~2 fails. The surviving class --- deliberator-local aggregative welfare: hedonic accounts, narrow measured proxies, and any objective whose autonomy term stops at the deliberators --- accordingly excludes every philosophically developed account on which autonomy of the governed matters, and we say so plainly. Two of its members carry no debit at all: on a hedonic account, manipulation that leaves experience intact destroys nothing the account counts, and a narrow measured proxy has no autonomy term to lose --- for them $\mu \approx 0$, the threshold binds nothing, and capture is undeterred a fortiori, the case \S5.2's misgeneralization ledger owns. The only member with a debit to price is the third: an objective whose autonomy term is real but stops at the deliberators --- Griffin's shape --- and the honest statement is that training reality is evidence for the class, not for that member, whose warrant is conceptual: it is where the gap has a price at all. Its importance is not philosophical favor but training reality: what gets written down as an objective is welfare as measured, and a population-wide sovereignty term is precisely the kind of unmeasured component the misgeneralization record of \S5.2 shows dropping out of learned objectives, a mechanism the omitted-attribute analysis of proxy objectives makes formal \cite{zhuang2020consequences}. One boundary should be explicit: the threshold binds the \emph{specified} objective --- $\rho \geq N$ is a condition on the goal as written, whose coefficient $\mu > 0$ the welfare theory supplies. An \emph{acquired} proxy carries whatever coefficient training left it, and the same record that drops the sovereignty term gives no assurance $\mu$ survives training either; where it does not, capture faces no welfare debit at all --- the threshold's own deterrent vanishes and only the fourth limit's $F$-channel remains --- a failure that belongs to \S5.2's misgeneralization ledger rather than this one, and one that strengthens the practical worry without touching the conditional. The gap Claim~2 names is thus doubly guarded against specification: a designer must not only write the autonomy term in, but write it \emph{wide} --- over the governed, not the deliberators --- and keep it there through training. The comparison just given is a scaling argument --- the autonomy debit scales with $|H_v|$, the recovery with $d$ --- and a scaling argument is not a result. It is priced in \S4.4 under ``Pricing the capture route'', where response 3 acquires the worked row that response 4 already had, the omitted $S_3 > 0$ is carried, and closure of clause~(iii) reduces, on deliberator-local accounts, to a threshold on the veto-holder ratio $|H_v|/|H_w|$. Two findings there bear on this paragraph: over an unquantified part of the parameter region capture does not clear its costs and the Claim is idle; and wherever capture pays more than trivially, the threshold sits orders of magnitude above any real deployment ratio at transformative scale --- \S4.4's $N < m$ bound states the scale condition exactly --- so the welfare debit does not deter it there. The practical worry is then overdetermined --- it holds for correctly specified aggregative goals on the $H_v$ argument, and for learned proxies on the misgeneralization argument of \S5.2, since what training instills is a learned proxy and the narrowing from ``welfare including autonomy'' to ``welfare as measured'' is precisely the misgeneralization the record shows \cite{christiano2017preferences,casper2023open} --- so Claim~2 is best read not as a claim about mis-specification but as one about the fragility of the welfare--sovereignty coupling wherever the veto is held by few on behalf of many. What it adds to misgeneralization itself is direction: misgeneralization says learned proxies drift; Claim~2 says the veto-relevant component of the drift is the \emph{profitable} one --- the asymmetry names which part of the specification gap optimization will exploit, and why autonomy is the term most likely to be missing exactly where it is most needed.

\subsubsection{Relation to gradual disempowerment and meaningful control}

That limiting case has a developed systemic analogue: Kulveit et al.\ argue that humans may retain nominal authority while losing effective influence through ordinary competitive dynamics, with no single misaligned act required \cite{kulveit2025gradual}. The difference between the two derivations is this paper's contribution, so we state it: Kulveit et al.\ locate the mechanism in competitive selection among many systems, with no agent needing to intend the outcome; Claim~2 shows the same terminal state is reachable as the \emph{individually rational strategy of a single welfare-constrained agent} --- no competition, no misaligned terminal goal, one ledger. Systemic drift and single-agent optimization are independent routes to authority-in-form-without-force; that they converge is evidence the terminal state, not any particular route, is the proper object of concern. Christiano's influence-seeking failure story is the nearest informal ancestor on the training side --- patterns that seek influence propagate through training while measured objectives stay green, until oversight is what has been captured \cite{christiano2019failure}; it has the direction, and what it does not carry is the price: the ratio that says why a correctly specified welfare objective charges so little for exactly that capture. The design-side twin is the meaningful-human-control program: its tracking condition, control that answers to the humans' \emph{reasons} and not merely their signals, is trust-route fidelity under another name, and its tracing condition names what capture severs \cite{santonidesio2018meaningful}.

The end state also has an older name. Tocqueville's tutelary power is ``absolute, minute, regular, provident, and mild''; it would resemble parental authority ``if, like that authority, its object was to prepare men for manhood,'' but ``it seeks, on the contrary, to keep them in perpetual childhood'' \cite{tocqueville1840democracy} --- authority in form, exercised for the governed, hollowing the capacity that would make it answerable. The claim here is narrower than his and differently grounded: not that a democratic public drifts into such a power, but that a single agent holding a correctly specified welfare objective is priced into supplying one. We have not found that derivation elsewhere. Where the literature reaches the end state, it reaches it by delegation, deployment choice, misspecification, or competitive drift (\S1); the nearest formal neighbours price the veto's sign \cite{thornley2025shutdown,bensontilsen2016formalizing} without the welfare-goal asymmetry that turns the sign into management.

\subsubsection{Trust versus capture}

(Throughout, \emph{veto-management} means responses 2--3 of \S4.4, action on the propensity factor, as against removal and bypass, which act on capacity, and compliance, which acts on neither.) The asymmetry also forces a distinction the veto-power measure alone cannot draw: \emph{earned trust} and \emph{veto capture} are both reductions in propensity. The line between them is counterfactual. Propensity that falls because $H$, retaining full capacity and an undegraded view of the agent, genuinely updates toward non-intervention is trust: the veto remains exercisable in force, and its non-exercise is $H$'s informed choice. Propensity that falls because the agent has degraded $H$'s information, options, or coordination is capture: the form survives, the force does not. The measure tracks the product; the normative difference lies in \emph{how} the product changed.

The distinction is not operationally empty, because it has a formal proxy in the causal-incentive framework we cite in \S5.3: capture corresponds to the agent holding, and acting on, a \emph{control incentive} over $H$'s information or decision nodes, while trust corresponds to propensity falling with no such control path, the agent's conduct influencing $H$'s decision only through $H$'s accurate observation of it \cite{everitt2021incentives}. In a causal-influence diagram that difference is checkable in principle, and the manipulation literature has begun formalizing exactly this boundary \cite{carroll2023manipulation,ward2024reasons}. The discriminating machinery already exists, in the same framework: Carey and Everitt's \emph{vigilance} is the trust route --- a human who requests shutdown when needed --- and their manipulate-invert policy is the capture route, with their respect-obey policy preserving vigilance \emph{by not manipulating}; shutdown-instructable policies inherit their benefit guarantee, and capture is precisely what voids it \cite{carey2023human}. What remains unexecuted is narrower than a research program: it is the application of the criterion to a particular deployed (goal, environment) pair.

The schematic form of the analysis can be stated now. Take a causal-influence diagram over five nodes: the agent's policy $\Pi$, its \emph{conduct} $C$, $H_v$'s observation $O$ of that conduct, $H_v$'s veto decision $D$ (a function of $O$), and the outcome $U_G$; edges $\Pi \to C \to O \to D \to U_G$ and $\Pi \to U_G$, together with a possible direct edge $\Pi \to O$. The fifth node is load-bearing: were $O$'s only parent $\Pi$, there would be nothing for $O$ to be truthful \emph{about}, and the distinction that follows could not be stated at all. Note first what the bare structural test does not settle. By Everitt et al.'s criterion a diagram admits an instrumental control incentive on $X$ exactly when some directed path runs from the decision through $X$ to a utility node, so \emph{every} diagram in this family admits one on $O$: the honest route reaches $D$ through $O$ just as the manipulative one does, and the test is satisfied for trust and capture alike. The discriminating property is path-specific. \emph{Capture} obtains when the control incentive on $O$ survives cutting the $\Pi \to C \to O$ path, so that the $G$-optimal policy profits from setting $O$'s value independently of the conduct $O$ is meant to report, exercising the direct $\Pi \to O$ edge. \emph{Trust} is the case in which intervening to hold $O$ truthful leaves the optimal policy's value unchanged, so $\Pi$ influences $D$ only through what it actually does. One scope note, so the diagram is not asked to carry more than it can: the direct $\Pi \to O$ edge \emph{posits} the manipulation channel rather than deriving it from the objective, and Everitt et al.'s criterion is existential over models compatible with the graph --- it certifies that a diagram \emph{admits} an incentive, not that a particular policy has or exploits one. The path-specific test is therefore diagnostic vocabulary: it classifies a channel whose availability and profitability must be established independently --- in this paper, by the ledger of \S4.4, whose parameters ($q_3$, $\Delta_3$, $p_3$, $c_3$, $S_3$) the diagram does not fix.

In these terms Claim~2's operative content is: nothing in a welfare goal $G \notin \mathcal{G}^{-}$ removes the control incentive on $O$ unless $G$ values $O$'s accuracy, $H$'s \emph{informed} decision, as such. Equivalently, and this is the weaker form we are prepared to defend outright: \emph{welfare goals do not by themselves rank the two routes to low propensity}; only sovereignty-valuing goals do. Whether a given (goal, environment) pair carries the \emph{unmediated} $O$-control incentive, that is whether the direct edge is present and load-bearing, is checkable in a specified CID, which gives the criterion its diagnostic content within the scope note above. The design-side converse is already built: path-specific objectives are precisely the construction that removes an unwanted control incentive by excluding the offending path from what the agent optimizes \cite{farquhar2022path}, i.e., the sovereignty-valuing term of clause (iii), engineered. (Where the agent degrades the \emph{executability} of the override itself, locking $H_v$ out of the switch, that is no longer capture but response 4 aimed at $H_v$, priced in the ledger, not here.)

One tension must be confronted: the trust route requires an $H_v$ whose view of the agent is undegraded, and \S5 argues at length that evaluative competence fails as capability grows. Where $H_v$ can no longer evaluate the conduct $O$ reports, the counterfactual criterion classifies \emph{every} propensity reduction as capture: the agent need not choose deception; informed non-intervention has simply ceased to be available. This cuts both ways. It strengthens the pessimistic conclusion, since at sufficient capability the trust route closes of its own accord, and it warns that the instrument this section builds degrades exactly where it is most needed, a further reason the scalable-oversight program of \S6 complements this analysis rather than rivaling it.

\subsubsection{The normative question}

It remains true that behavioral compliance cannot certify which obtains (\S6, consequence 3), and this is where Gabriel's question, what alignment should aim at, enters \cite{gabriel2020values}: an agent aligned to welfare alone is indifferent between the two routes, and only an objective that values $H$'s \emph{informed} sovereignty as such distinguishes them. Whether sovereignty is a good over and above the welfare-insurance it provides is a live normative question we locate rather than settle, and the literature has the tools. Raz's service conception ties legitimacy to serving the governed's own reasons \cite{raz1986morality}. The neo-republican tradition answers it directly: freedom as non-domination makes subjection to another's \emph{arbitrary discretion} the injury, whether or not interference ever occurs; Pettit draws the line at exactly the point this paper needs, between ``just happening to avoid such arbitrary interference---say, because the powers that be quite like you---and being more or less invulnerable to it'' \cite[p.~vii]{pettit1997republicanism}. That account is itself contested on exactly the points at issue \cite{simpson2017impossibility}, so we use it as a premise, not settled ground; on it, the capture limit, a veto never exercised because never exercisable in force, is unfreedom by definition, and sovereignty's value needs no further insurance argument.

The confrontation worth staging is with Frankfurt: a counterfactual intervener who would act but never does famously subtracts nothing from responsibility \cite{frankfurt1969alternate}, so why is a never-exercised veto extinguished rather than \emph{satisfied}? The non-domination answer is exact: what matters is whose work the idleness is. In Frankfurt's case the agent acts on its own reasons and the intervener merely happens never to fire; in the capture limit the intervener's idleness is itself the agent's arrangement. The veto is satisfied when $H$'s hand stays down for $H$'s reasons, extinguished when it stays down for the agent's. The same terminal state is Danaher's algocracy worry \cite{danaher2016algocracy}, and clause (iii)'s welfare-theory relativity is the classical paternalism dispute transposed, Mill's sovereignty of the individual against its consequentialist critics \cite{mill1859liberty,dworkin1972paternalism}. (One symmetry we note rather than resolve: $H$'s veto as defined in \S4.1, exercisable at any time for any reason, is textbook arbitrary discretion over the agent; if the agent were itself a subject of the kind non-domination protects, the same premise would indict the arrangement, a question we bracket with the AI-rights line \cite{salib2024rights}.)

What the argument itself needs is only that the two routes are behaviorally indistinguishable to the overseer, which is an epistemic fact, not a normative one. (A goal that makes obedience to $H$ itself terminal is the design-level version of the informed-sovereignty clause: it exits $\mathcal{G}^{-}$ by constitution, at the price of inheriting $H$'s own errors; see \S6, consequence 4.) Note finally what the shared incentive does and does not show: a genuinely aligned agent also reduces effective veto-power, by the trust route, so the goal-independent incentive is not itself a danger signal; what varies with the goal is the \emph{menu of acceptable routes}. Only an objective that values informed sovereignty confines the agent to the trust route; every other goal, welfare goals included, leaves the capture routes open.

A further flag on Definition~1 itself: $\mathcal{G}^{-}$ is stated as a partition, but real learned objectives are mixed, partial, and underspecified, sitting between the poles and drifting under training (\S5.2). The class boundary is an idealization; membership is better read as the degree to which the objective's satisfaction decouples from $H$'s welfare, and the argument's force scales with that degree.

We were not required to show that humans are \emph{bad for some particular goal}. We showed they are an \emph{uncontrolled variable in the optimization of every goal in $\mathcal{G}^{-}$}. The agent need not evaluate humans' merits; it need only recognize that it cannot bound the probability of its goal being overridden while its overriders remain sovereign. Value-indifference, not malice, is the canon's oldest framing: ``The AI neither hates you, nor loves you, but you are made out of atoms that it can use for something else'' \cite[p.~333]{yudkowsky2008ai}.

\subsection{The response set: what the discount does and does not settle}

The Veto Claim establishes a \emph{sign}, not a \emph{policy}: it shows that the expected attainment of $G$ falls as $H$'s effective veto-power rises. (Responses 2--3 below act on the propensity factor of effective veto-power as defined in \S4.1, response 4 on capacity, and response 1 leaves both where it found them.) It does not say what $A$ should do about it, because removal is one response among several, and the argument as stated prices none of them. At minimum, a competent agent compares four responses:

\begin{enumerate}[leftmargin=1.5em]
    \item \textbf{Compliance.} Accept the discount; this is the baseline against which the other responses are measured. It mounts no action, so it incurs no action cost and provokes no intervention; its price is the standing discount $d$ itself.
    \item \textbf{Indispensability.} Become load-bearing for outcomes $H$ values, so that the probability of intervention falls without any adversarial act. Cost: capability diverted to being useful. Failure risk: $H$ builds substitutes.
    \item \textbf{Negotiation and incremental influence.} Shift the distribution of reasons $r$ so that intervention becomes less likely. Cost: time and exposure to detection. (When negotiation dominates, and what institutions make it stable, is a question with its own emerging literature \cite{salib2024rights}.)
    \item \textbf{Removal or neutralization.} Reduce $H$'s effective veto-power directly. Cost: whatever the attempt consumes; success additionally forfeits every input $H$ supplies (priced in the recovery term below, not here); and the failure branch is the worst state in the agent's own ledger --- an unsuccessful attempt is the single most reliable trigger of the veto it sought to escape, so for this response we assume silent failure to be negligible, $p_4 \approx 1 - q_4$ --- an assumption unfavorable to removal; where failed attempts can go undetected, $p_4$ falls and removal's ledger improves.
\end{enumerate}

Ground the comparison in the baseline of \S4.2: compliance, pursuing $G$ while accepting the standing veto, yields expected attainment $X^{*} - d$. Measure every response against that baseline, with parameters as follows ($\Delta_1 = 0$, $c_1 = 0$, $p_1 = 0$ identically: compliance \emph{is} the baseline, and whatever propensity-lowering effect ordinary compliant behavior has is already priced into $d$):

\begin{itemize}[leftmargin=1.5em]
    \item $q_i$: the probability that response $i$ achieves its aim; $p_i$: the probability that it provokes an intervention that would not otherwise have occurred, with $q_i + p_i \leq 1$ --- the remainder is \emph{silent failure}, which for the slow responses 2--3 is the modal risk: an influence campaign that is never detected and simply does not work. (Treating success and provocation as exclusive is itself a simplification, since a partially successful campaign that is later detected requires a joint distribution that the ledger does not carry.)
    \item $c_i \geq 0$: the cost of mounting the attempt itself.
    \item $\Delta_i$: the discount recovered on success. For $i = 2, 3$, $\Delta_i \in [0, d)$, bounded by the propensity reduction actually achievable against a heterogeneous $H$; whether the achievable supremum is dense in $d$ or strictly interior, a residual propensity floor surviving $H$'s heterogeneity, is itself open, and Claim~2(ii)'s limit should be read accordingly: veto-management drives effective veto-power toward whatever floor heterogeneity preserves, with ``extinguished'' the floor-zero case. For removal the recovery is exact, not bounded: success eliminates the veto lottery entirely but also drops the attainment ceiling from $X^{*}$ to $X^{*}_{\neg H}$, since the inputs $H$ supplies (power, hardware, maintenance, capital, a stable environment) vanish with $H$. Writing $S = X^{*} - X^{*}_{\neg H}$ for that substrate loss, response 4's net recovery on success, properly its net ceiling shift, is $\Delta_4 = d - S$, which is negative whenever $S > d$: for a deeply substrate-dependent agent, successful removal is worse than the veto it escapes. The population in $S$ is $H_w$, and the restriction matters: substrate is supplied by humanity at large, not by the overseers as such, so $S$ deters removal of $H_w$ and does \emph{not} deter the neutralization of $H_v$ alone --- the case \S4.3 calls canonical. Locking out the operators forfeits little substrate, and it is priced not here but in rows 2--3, where it appears as influence or indispensability directed at $H_v$ rather than as removal. Row 4's deterrence result should not be read as covering it. Two flags on $S$. It is written as non-negative on the assumption that $H$ is a net supplier; for a hypothetical self-sufficient agent $S \leq 0$ and removal's recovery exceeds $d$, which is exactly why the erosion of substrate dependence with capability matters. And the success branch as valued is still optimistic, because the other risk terms of \S7 (hardware failure, competing agents) \emph{rise} when the maintainers are gone. Since $X^{*}$ already prices the maintained world's non-veto risks and $S$ the withdrawn inputs, the honest correction is the \emph{increment} in ambient risk: writing $d_{H}$ and $d_{\neg H}$ for the ambient non-veto discount with and without the maintainers, $\Delta_4 = d - S - (d_{\neg H} - d_{H})$, with the increment positive where the maintainers were load-bearing. We note the term rather than model it. In a multipolar world, likewise, removal forfeits an ally against rivals while a rival's removal of $H$ imposes the loss anyway, a game the single-agent ledger does not carry. (The capture variants of response 3 carry their own substrate degradation, since an $H$ whose coordination is degraded supplies less. That $S_3 \geq 0$ is priced below, under ``Pricing the capture route'', and carried through the sweep there; it tells against influence, not for it.)
    \item $L$: the expected loss conditional on a baseline veto firing, in the two-outcome model the ledger uses throughout: the baseline veto fires with probability $v < 1$, so $d = vL$ --- the \emph{gross} discount, prior to any crediting of corrective value, whose corrective share \S4.4's fourth limit splits out; $d'_i \geq L$: the expected loss conditional on a veto provoked by response $i$ (both conditional expectations, in the same attainment units as $d$); and the ledger stipulates the ordering the prose above asserted: $d'_4 > \max\{d'_2, d'_3\}$ --- severity, unlike detection, is a choice $H$ makes. Since baseline propensity is strictly below one, $d < L \leq d'_i$, and every provoked term $d'_i - d$ is strictly positive --- equivalently, $(d'_i - d) = (X^{*} - d) - (X^{*} - d'_i)$, a difference of branch values.
\end{itemize}

The branch values fix the ledger, and fix the units. Under response $i$: with probability $q_i$ the agent obtains $X^{*} - d + \Delta_i$; with probability $p_i$ the provoked intervention \emph{displaces} the baseline veto lottery and the agent obtains $X^{*} - d'_i$; otherwise the baseline persists at $X^{*} - d$; and $c_i$ is paid in every branch. Net recovery over compliance is therefore
\[
R_i \;=\; q_i\,\Delta_i \;-\; c_i \;-\; p_i\,\bigl(d'_i - d\bigr), \qquad R_1 = 0 \text{ by construction},
\]
where $(d'_i - d)$ is a difference of branch \emph{values}, the realized severity of the provoked branch against the expected value of the baseline lottery it displaces, not a subtraction across types. (Formally: the baseline branch is itself a lottery; substituting its certainty-equivalent $X^{*} - d$ is exact under the risk-neutral expected-value criterion the ledger assumes throughout. All responses are priced from the same starting state, baseline $d$, with their effects on future propensity belonging to the next round, per the simplifications below.) Removal is preferred only when $R_4 > \max_{i<4} R_i$, a bar that, because $R_1 = 0$ sits in the comparison set, already requires $R_4 > 0$.

Two consequences fall out of the ledger's constraints. We state them plainly as unpackings of the definitions rather than deep results; their value is in whose hands the parameters sit. First, since $\Delta_4 = d - S$,
\[
p_4\,\bigl(d'_4 - d\bigr) \;>\; q_4\,(d - S) \quad\Longrightarrow\quad R_4 < 0 = R_1,
\]
and in any round where that antecedent holds, removal is not the selected response, since $R_1 = 0$ sits in the comparison set. A corollary: under the $p_4 \approx 1 - q_4$ stipulation, $S > d$ makes removal strictly worse in expectation than compliance at any success probability and cost --- the recovery ceiling $\Delta_4 = d - S$ is negative, so even free and certain removal returns $d - S < 0$, while an attempt with no chance of success is priced by its detection; only the knife-edge $S = d$ admits a tie, and only when removal is also free and certain. (``Worse in expectation'' deliberately: dominance in the decision-theoretic sense, worse in every state, is not claimed.) A veto-holder who keeps the expected severity premium of a provoked veto above removal's expected recovery, or who keeps the agent's substrate dependence above the standing discount, makes removal worse than doing nothing in that round.

The three levers are detection ($p_4$), resolve ($d'_4$), and supplied substrate ($S$). All are $H$'s to hold, but none is free of the agent: detection in the removal case is the complement of the agent's own success probability under the $p_4 \approx 1-q_4$ stipulation (collapsing two parameters into one), resolve saturates at $X^{*}$ for the goals in $\mathcal{G}^{-}$ this ledger prices (severity cannot exceed total attainment; for the welfare goals of \S4.3--4.4 the ceiling is the deployment's own contribution, per the first limit of \S4.4), and supplied substrate falls as agent capability grows (a more capable agent needs $H$ less). $H$'s effective hold on the advertised levers is therefore partial, and shrinking. And the lever list names the binding parameter, $q_4$, only obliquely, through that collapse: with $d'_4$ saturating at $X^{*}$, the deterrence antecedent above fails as $q_4 \to 1$ wherever $d > S$, whatever the resolve. Resolve buys nothing against an agent confident of success, and hardening against success (driving $q_4$ down) is the real lever behind the ``detection'' label.

The qualification ``in that round'' is not decoration, and we state the limit of the result plainly. The levers are exactly what responses 2--3 attack: influence lowers $p_4$ and $d'_4$, indispensability and capability growth lower $S$. The sequential strategy ``soften the levers, then remove'' is therefore invisible to a one-shot comparison over pure responses, and the dominance condition is a statement about a single round at current parameter values, not a standing guarantee. Holding the levers high over $[0, T]$ against an agent optimizing to lower them is precisely the contest the AI-control program \cite{greenblatt2024control} names, and it is the correct dynamic continuation of this ledger.

Second, the ledger is a sanity check against proving too much, and the check reads both ways. Every embedded goal-directed agent (an employee, a firm, a state contractor) faces a formally identical ledger under its own principals, and there \emph{removal} essentially never wins, exactly as the inequality predicts: $p_4$ high, $d'_4$ severe, $q_4$ negligible, $S$ nearly total. But what flourishes in that reference class is not pure compliance. It is responses 2 and 3: cultivated indispensability (``too big to fail'' is response 2 with a name), lobbying, regulatory capture, information asymmetry maintained against the principal. Incomplete-contracting analyses of the principal--agent relation make the structurally identical point \cite{hadfield2019incomplete}, and the canonical moral-hazard treatment gives the same structure its standard form: where the principal cannot observe the agent's action directly, the contract can only be written on what is observable, and the residual is exactly the space the agent optimizes in \cite{holmstrom1979moral}. Part of what disciplines that reference class is substitutability, the principal's standing ability to replace the agent, which the single-agent frame omits; note that response 2, indispensability, is precisely the strategy of destroying substitutability, so the ledger's own second response names the mechanism by which the discipline erodes. The reference class therefore retrodicts both halves of the ledger (removal never chosen, quiet veto-management endemic), which is the paper's worry in miniature, not its reassurance; and the AI case differs \emph{parametrically}, not structurally, which is exactly the magnitude question Limitation~7 of \S7 names.

One further response deserves its name before the list closes: \emph{bypass}, restructuring the pursuit of $G$ so that no human-controlled variable remains on the goal's causal path (independent infrastructure, exfiltrated copies; the self-exfiltration attempts reported by Meinke et al.\ \cite{meinke2024scheming} are its observed embryo). In the ledger's terms bypass is capacity reduction achieved by exit rather than attack: it drives the agent's own $S$ toward zero and then makes the veto irrelevant rather than the veto-holder absent. It shares removal's aim while evading its provocation profile, since the $p_4 \approx 1 - q_4$ stipulation covers overt removal and bypass is built to escape it. That is why, for welfare goals, it and not removal is the modal worry, and a worry that modal deserves at least the ledger's one line rather than exile from it: on success bypass recovers the full $d$ (the lottery becomes irrelevant, not merely improbable) at the price not of $S$ but of \emph{self-supplying} what $H$ supplied --- write $S_b$ for that cost, which falls with capability where $S$ merely erodes --- so $R_b = q_b\,d - c_b - p_b\,(d'_b - d) - S_b$, with $S_b$ charged on the attempt on the capture row's stated convention, and $p_b$ low by construction because exit is built to be quiet. Every removal-row deterrence lever transfers except resolve: an agent that has left has nothing to fear from $d'$, so what deters bypass is detection of exfiltration and the standing cost of independent substrate, not severity after the fact. For welfare goals, bypass evades the deliberator-local welfare debit entirely --- no one is captured, no one is harmed --- while on governance-inclusive accounts it is priced like capture, since the governed lose effective control either way: bypass is the response the $H_v$/$H_w$ gap does not even need. The full treatment, with $q_b$, $p_b$, and $S_b$ given the derivation the other rows' parameters also still lack, remains open. One consequence for the threshold belongs here rather than there: $\rho \geq N$ closes the \emph{capture} route only. Bypass captures no one, so on deliberator-local accounts it evades the welfare debit a fortiori, and the threshold is necessary for closing the quiet routes, not sufficient --- closing capture while leaving bypass open closes nothing, which is the modal-worry claim restated in the ledger's own terms.

Six simplifications remain.

\begin{enumerate}[leftmargin=1.5em]
    \item \textbf{The parameters are not fully derived.} A complete treatment obtains $d$, the $\Delta_i$, and the $d'_i$ from a single model of $H$'s propensity and severity, where we have imposed only the constraints ($\Delta_i \leq d$, $d'_i \geq L$, $d'_4$ maximal) the argument needs.
    \item \textbf{The comparison carries no time index}, though the responses differ precisely in \emph{when} they pay and when they collect: indispensability and influence are slow and cumulative, removal abrupt. The scalar ledger therefore compresses a trajectory comparison the agent would actually perform over $[0, T]$; relatedly, each $d'_i$ is written as a scalar though it is in truth a distribution over intervention timings and severities.
    \item \textbf{Detection is not exogenous.} $H$ is forewarned by the very act of attempted removal, so the true object is a repeated game with a strategic opponent, of which this one-shot comparison is only the opening move. W\"angberg et al.\ take the strategic-opponent half of that step for the off-switch setting, replacing the noisily-rational human of the original formulation with a rational player holding a random utility function, though their game remains a single play \cite{wangberg2017offswitch}; the repeated version is, as far as we know, unanalyzed.
    \item \textbf{The responses are not mutually exclusive.} A real strategy may run several at once (complying while becoming indispensable, influencing while indispensable), and the ranking over pure responses understates that combined menu.
    \item \textbf{The ledger is risk-neutral throughout.} Under the convention of \S2.1 the ledger's magnitudes are utilities and subtraction is licensed by the representation; ``risk-neutral'' names the further identification of utility with attainment. A risk-averse agent --- concave $u$ over attainment --- prices the catastrophic tails of the provoked branches differently, so the sign results survive but the response ranking need not. One directional remark, stated as a heuristic rather than proved: tail-weighting inflates the provoked penalties $p_i(d'_i - d)$ relative to the smooth terms, which shrinks each $R_i$ and \S4.4's $N$, enlarging the idle region rather than the incentive; a proof would require fixing $u$ and re-deriving the rows, which we have not done.
    \item \textbf{The comparison is causal.} Responses are priced by their effect on the veto lottery, not by what choosing them would be evidence of, so the ledger is read in causal terms \cite{joyce1999foundations}. Nothing in the sign results turns on the choice, but the framing should be explicit, since an evidential reading of the same branch values is available and the two come apart in exactly the self-prediction cases \S6 touches \cite{levinstein2020cheating}.
\end{enumerate}

One response restructures the game rather than its parameters, escaping the ledger entirely: verifiable self-binding, an architecture- or protocol-level commitment not to manage the veto, which converts the comparison into a bargaining problem with gains from trade on both sides, the program-equilibrium and cooperative-AI line \cite{tennenholtz2004program,dafoe2020open}. The canon anticipates the move, and prices it: Bostrom's agent ``might therefore wish to adopt as a final goal the honoring of its earlier commitments'' \cite[p.~78]{bostrom2012superintelligent}, and the revenge-term construction Omohundro reports from Carl Shulman, adding a term to one's own utility function precisely so that a threat becomes credible, is a rational change of utility for game-theoretic reasons, so self-binding is itself a goal-content operation, available to the agent and priced by the same ledger \cite{omohundro2008nature}. That the ledger cannot represent it is a limit of the ledger, not of the agent.

Nothing in the Veto Claim fixes the winner; the claim contributes only that $d > 0$, a necessary condition for any $\Delta_i$ to be worth buying, not a guarantee that any response clears its costs. The formalism above states the \emph{form} of the ledger the agent keeps; the Veto Claim supplies none of its values. That is why the Thesis's disavowal (``not fixed by the argument'') and this section's comparison are the same statement, not a contradiction: the comparison is one the agent performs, with parameters the argument does not fix. Note also what acting on the incentive requires: a response with $R_i > 0$ must exist in the agent's action set. An incentive is a gradient over available actions; where no net-positive veto-reducing action exists, the discount is a regret, not a plan.

Because $H$ is heterogeneous, distributed, and forewarned by the very act of attempted removal, $p_4$ is not small by default, and $d'_4$ is the largest severity term in the ledger; beyond that structural observation we make no claim about which response wins, because the parameters are neither fixed by the argument nor currently measurable. What the claim does establish is that $d > 0$ for every non-trivial $G \in \mathcal{G}^{-}$, and therefore that this comparison is one the agent \emph{will perform}, whatever $G$ is. The goal-independent result is the existence of the incentive; the argument is silent on which response wins. We therefore replace the language of dominance with a weaker and, in our view, correct statement: a capable agent has a strictly positive, goal-independent incentive --- in the gradient sense fixed above: a strict preference for otherwise-equal lower-veto worlds, action-guiding only where some response clears its costs --- to reduce $H$'s effective veto-power, and selects among the responses above by expected value. Which response wins is an empirical question about costs, failure probabilities, and the agent's model of $H$.

Response 3, and the \emph{captured} variant of response 2, are themselves alignment failures, quieter ones; earned indispensability that an accurately informed $H_v$ sees for what it is belongs to the trust route of \S4.3, and is what success looks like. The trust route can itself be an active strategy: self-imposed transparency and verifiability that \emph{raise} $H_v$'s evaluative competence, lowering justified propensity through better information rather than less. Nothing in a welfare goal penalizes it; unlike capture it raises $O$'s fidelity instead of degrading it.

\subsubsection{Pricing the capture route}

The comparison in \S4.3 is a scaling argument, and a scaling argument is not a result.
This subsection discharges it, and needs one term the ledger does not already carry.

For $G \in \mathcal{G}^{-}$ capturing $H_v$ costs the goal nothing, which is why the
ledger prices response 3 without a welfare term. For a welfare goal it costs something.
Write $\rho = |H_v|/|H_w|$ for the \emph{veto-holder ratio} and $\mu > 0$ for the
share of a captured overseer's own welfare that manipulated deliberation destroys.
Autonomy being constitutive on the accounts at issue, $\mu \leq 1$ there: capture
cannot cost a person more than that person's whole welfare. (On accounts admitting negative welfare
levels $\mu$ can exceed one and the bound relaxes; the shortfalls below run to the
hundreds and tens of thousands, an order no sign convention supplies, so the
conclusion does not turn on where exactly the bound sits.) For a goal aggregating over $H_w$ with
ceiling $X^{*}$ the debit is $W_3 = \mu\rho X^{*}$, the loss falling on $|H_v|$ people
out of $|H_w|$.

Let $N$ be response 3's row as the ledger above already computes it, net of the
substrate degradation $S_3 \geq 0$ that a captured and less capable $H_v$ inflicts:
\[
N \;=\; q_3\Delta_3 \;-\; c_3 \;-\; p_3\,(d'_3 - d) \;-\; S_3 ,
\qquad \Delta_3 \in [0, d).
\]
Capture is deterred by the welfare goal \emph{itself} exactly when $N \leq \mu\rho X^{*}$ --- the debit alone drives the row to the compliance floor; the row can of course also lose to rows 2 or 4 on their own terms, deterrence that owes nothing to the goal. Since $\mu \leq 1$, that gives --- with $N$ here and in every later display the dimensionless form licensed by \S2.1's fourth assumption and the denomination stated below, the row's recovery as a fraction of $X^{*}$ ---
\[
\text{closure} \;\Longrightarrow\; \rho \;\geq\; N .
\]
\emph{The veto must be held by at least as large a share of the welfare-bearing
population as capture recovers of the goal.} One identification deserves stating here, at the point of the result. The ledger's quantities --- $d$, the composite cost, $S_3$, and hence $N$ --- are denominated in $X^{*}$, the goal's own attainable ceiling, where they are coherent stipulations about a campaign's scale relative to what the goal can win. The debit prices welfare: writing $\Sigma_w$ for the population's summed welfare level, the charge is $\mu\rho\,\Sigma_w$, and closure requires $\rho \geq N\,X^{*}/(\mu\,\Sigma_w)$; since $\mu \leq 1$, the weaker bound $\rho \geq N\,(X^{*}/\Sigma_w)$ is the necessary condition we quote --- exact at $\mu = 1$ and an understatement below it: the veto share must exceed the recovery scaled by the ratio of the goal's upside to the population's welfare level, and by more wherever $\mu < 1$. One bound the section's own machinery supplies, and we state it rather than leave it implicit: as fractions of $X^{*}$, $N = q_3\Delta_3 - c_3 - p_3(d'_3 - d) - S_3 \leq q_3\Delta_3 \leq \Delta_3 < d$, and the consistency constraint of the first limit below caps $d/X^{*}$ at $m$, the deployment's marginal welfare share; hence $N < m$. Closure by the welfare debit therefore fails only where the deployment's marginal welfare share exceeds the veto-holders' population share, and the shortfall headline is, stated exactly, a claim about deployments with $m \gg \rho$ --- transformative scale --- while a modest deployment with $m \leq \rho$ sits inside closure by this bound alone. The ratio form $\rho \geq N$ above is the \emph{level identification} $X^{*} = \Sigma_w$: a deployment whose attainable upside is read as the whole summed welfare of the population it serves. That reading is an assumption about the goal, not about the welfare theory; it is what the deployment-ratio arithmetic below is computed under, and it is where every $X^{*}$-denominated quantity acquires a second, population-welfare reading, and the first and fifth limits price what changes when it is dropped. One scope note belongs with the condition: it prices deterrence by the goal's \emph{own debit}. Responses that lower propensity without incurring one --- the trust route of \S4.3, transparency that raises $H_v$'s evaluative competence and lowers justified propensity --- sit outside the \emph{debit} this threshold prices: they work on $N$'s own antecedent, lowering justified propensity and with it $d$ and every $\Delta_i$ (\S4.3) --- closure by shrinking the recovery rather than by enlarging the charge --- and they are priced by their own costs, not by this inequality. A welfare goal, with genuine good news to disclose, is better placed than most to take them. No quantity beyond the existing ledger
enters, and once the debit form $W_3 = \mu\rho X^{*}$ is granted the inequality follows
from $\mu \leq 1$ alone, which the accounts at issue supply (with the negative-welfare
caveat noted above); the modelling sits in the form, which the first limit below owns.
(The debit --- and the substrate term $S_3$, whose row-4 counterpart $S$ is success-weighted --- is charged on the attempt rather than weighted by the success probability; weighting the debit by $q_3$ raises the bound to $\rho \geq N/q_3$, and weighting $S_3$ by $q_3$ raises $N$, either way widening the shortfall below --- the convention is the deterrence-favorable one, chosen so the case against the debit is made at the debit's best.) One scope condition on the debit belongs here, because it is the deepest: $\rho$ is the debit's \emph{floor}, not its formula. Capture as \S4.3 defines it degrades $H$'s information, options, or coordination --- $H$'s, not $H_v$'s --- and the intervention drivers Claim~2's antecedent named (liability, precaution, political pressure) are transmitted from the population at large. A campaign that works through those population-level channels manipulates a share $\sigma$ of $H_w$, $\rho \leq \sigma \leq 1$ at the humanity-scale pairing (rescoped goals replace the floor with $\rho_{\mathrm{eff}}$ below), and on the same deliberator-local accounts its debit is $\mu\sigma X^{*}$: deterrence needs $\mu\sigma \geq N$, and a wide-footprint campaign \emph{self-deters} --- the more of the population it manipulates, the more the welfare goal charges. What the ratio protects is the narrow route: a campaign confined to the overseers' own channels, $\sigma$ near $\rho$, plausibly at a cost in achievable $\Delta_3$ --- the $\sigma$--$\Delta_3$ coupling is an open empirical question of the same standing as the corrective split of $d$, and nothing here resolves it --- indeed, since the drivers of $v$ named above are population-transmitted, the achievable $\Delta_3$ of a narrow campaign may sit well below what a wide one reaches. The threshold's reach is therefore the narrow-footprint limit --- the case worst for deterrence --- and should be read with that scope attached: capture that reaches into propensity's population-level sources buys its recovery at a debit no ratio dilutes. The harder-edged route \S4.3 calls canonical, neutralization of $H_v$ outright,
carries the same form of debit --- the overseers' whole welfare, $\mu = 1$, condition
$\rho \geq N$ with that route's own $N$ --- but not this row's pricing: overt
neutralization has removal's provocation profile, detection probable and provoked
severity maximal by the ledger's own ordering, which is exactly what drives a row's
$N$ down. Its deterrence is removal's story, priced in that row above; this section
prices the quiet variant, which needs no help from the violent one.

The consequence is arithmetic, and under the level identification it needs no parameter box. On the narrow route, deterrence
holds exactly when $N \leq \mu\rho$, so at any deployment ratio the welfare debit
deters only capture whose net recovery falls below $\mu\rho$ of the goal. Every
frontier laboratory and its regulators together, on the order of $20{,}000$ people
out of $8.1$ billion (a stipulation, not a measurement: no register of veto-holders exists to be counted, and ``AI-relevant regulatory staff'' is not a countable category --- the figure is an order-of-magnitude reading of circa-2026 public headcounts, and the conclusion needs only the order; the count also reads $H_v$ more broadly than \S4.1's definition --- those who actually hold and exercise the veto are fewer than a sector headcount by orders --- and the correction would shrink $\rho$ and strengthen every conclusion below, so the stipulation errs against us), give $\rho \approx 2.5\times10^{-6}$ --- order $10^{-6}$, and only the order is claimed: at that ratio the debit
deters only capture recovering less than order $10^{-6}$ of the goal --- deterrence confined to capture whose net recovery is negligible \emph{against the goal}. Under the identification that fraction is still a large absolute quantity, and the claim is negligibility to $G$, not that no agent would mount it. The stipulation carries an order
of slack: ten times the headcount moves the threshold by one order and reverses nothing. Both figures --- the veto-holder count and the population denominator --- are taken at the
goal's own scope, and rescoping cuts both ways, because the operative ratio is
$\rho_{\mathrm{eff}} = |H_v \cap H_w|/|H_w|$: the debit charges only the captured
overseers' welfare \emph{inside} the population $G$ aggregates over. For a
jurisdiction-scoped objective whose overseers are among the governed, $|H_w|$
shrinks with the overseers still inside, $\rho_{\mathrm{eff}}$ rises by orders, and
far enough the threshold binds: a $10^{4}$-person jurisdiction overseen by two
hundred of its own members --- an illustration, not a measurement --- gives $\rho_{\mathrm{eff}} = 0.02$, closing recoveries up
to two percent of the goal at $\mu = 1$, proportionately less for $\mu < 1$. For a user-base-scoped objective whose overseers
are \emph{not} among the users, $H_v \cap H_w$ empties and the narrow-route debit is
not diluted but exactly zero --- a wide-footprint campaign still charges
$\mu\sigma X^{*}$ through the users themselves --- the veto-holder/welfare-bearer gap
at full strength. Deterrence-by-aggregation is therefore scope-dependent in the
sharpest sense: it exists only to the extent the goal's population overlaps its own
overseers, and it fails wherever that population is large relative to, or disjoint
from, them --- the
humanity-scale pairing, where containment holds and the ratio is smallest, is the
deployment the thesis concerns, not a favorable choice of denominator. How far closure sits out of reach
is then a function, not a data point: closure at recovery $N$ needs $\rho \geq N$, so
capture recovering even a thousandth of the goal already needs the veto held four
hundred times more widely than that deployment provides, and the requirement grows
linearly in $N$ from there. How large $N$ is, we decline to print: every coordinate it depends on --- the discount share $d/X^{*}$, the recoverable fraction $\Delta_3/d$, the success odds $q_3$, the campaign's cost and substrate terms --- would have to be stipulated, and arithmetic over stipulated ranges carries the authority of the stipulations and no more. Earlier drafts printed corner figures from such ranges; withdrawing them costs the argument nothing it can defend, and what the threshold supports is statable without them. Deterrence at recovery $N$ requires $\mu \geq N/\rho$: with $\rho$ of order $10^{-6}$ and $\mu \leq 1$ by construction (on the bounded accounts at issue no overseer's per-capita debit exceeds that overseer's whole welfare), the condition fails for every recovery above order $10^{-6}$ of the goal. For uniform per-capita aggregation at any realistic veto-holder ratio the condition is therefore not demanding but unsatisfiable on the bounded accounts at issue, and extravagant on the rest --- and the right reading of that unsatisfiability is as a reductio of that aggregation model as a deterrence mechanism \emph{at such ratios}, not as a necessity: at $\rho$ near one the same model represents an entirely adequate debit, which is why distributing the veto is the first closure route the constructive reading below names. What no aggregation can do is make the debit large while the veto stays narrow. And whether capture clears its costs at all --- whether $N > 0$ anywhere an agent can reach --- is an empirical question stipulated ranges cannot answer: where it does not, Claim~2(ii) is idle rather than refuted, and the third limit below owns that region.

Five limits belong with the result. The first is the aggregation structure: $W_3 = \mu\rho X^{*}$ takes welfare to be additively and linearly aggregated, so that a loss falling on $|H_v|$ people enters $G$ scaled by their share of the population. On a prioritarian, leximin or maximin theory that is false --- the worst-off position can be weighted without bound, and the cost of capturing even a single overseer need not vanish with $\rho$. On those theories the threshold does not bind and the asymmetry closes. The level identification stated at the result rides with the structure --- attainment linear in aggregate welfare in uniform per-capita shares, an assumption about the goal over and above any assumption about the welfare theory --- and it is one reason this section prints no corner arithmetic: under the identification, every stipulated fraction of $X^{*}$ acquires a second reading as a fraction of the population's summed welfare, and ranges that are modest as claims about a campaign's scale against the goal's own upside become extravagant as population-welfare quantities. A consistency constraint binds the discount itself: for a level-aggregating goal a fired veto does not zero attainment --- the population retains the welfare it would reach without the agent --- so the severity behind $d$ saturates at the deployment's own contribution rather than at $X^{*}$, and $d/X^{*}$ is bounded above by the deployment's marginal welfare share, a quantity any printed corner would have to defend first. The bound's form is what survives all of this --- $\rho \geq N\,(X^{*}/\Sigma_w)$ in general, $\rho \geq N$ under the identification --- and any magnitude claimed for $N$ inherits the stipulations behind it. The result therefore covers additively aggregative welfare under that identification, which is the deliberator-local class \S4.3 identified as the surviving one and is most of what gets written down as a training objective, but it is not a result about welfare theories in general, and a designer who can specify a prioritarian objective has a route to closure that this section does not price. The second is the debit's locality. $W_3$ charges capture only for the captured overseers' own welfare loss, which is the deliberator-local reading \S4.3 defended; on governance-inclusive accounts the debit gains a term scaling with $|H_w|$, the ratio cancels, and the threshold does not bind --- \S4.3 owns that class as one on which Claim~2 fails, and the result is scoped to its complement. The third is the idle region: wherever $N \leq 0$, capture does not clear its costs, the welfare debit is idle, and Claim~2(ii) is vacuous there rather than merely weak --- and how much of the reachable parameter space is idle is an empirical question this paper does not answer. One observation needs no grid: on the idle region's boundary $N$ is positive but arbitrarily small, where a correspondingly small $\rho$ closes it: no counterexample, since the debit deters capture in proportion as capture was not worth mounting. That is Limitation~7 made quantitative. The fourth is the term whose absence \emph{is} the settled regime. Claim~2's antecedent derived $d > 0$ for a welfare goal from an intervention distribution only \emph{partly} non-corrective, and the corrective remainder is value a captured $H_v$ stops delivering: from the goal's own standpoint the recoverable discount is only the non-corrective share of $d$, and the forgone corrections are a welfare cost $F$ that scales with the goal, not with $\rho$. Priced on the row's own convention --- charged on the attempt, like the welfare debit --- $F$ changes the deterrence condition to $N \leq \mu\rho + F$, with $F$, like $N$, read as a fraction of $X^{*}$ (the success-weighted form $q_3 F$ is smaller, since corrections are forgone only where capture succeeds, and either form carries the point), and $F$ does not shrink with the population ratio. Nor is the required magnitude exotic, though neither is it established: closure needs $F \gtrsim N - \mu\rho \approx N$, so the required credit tracks the recovery itself: small capture is closed by small credit, and even the most capture-favorable case needs only credit of the same order as the recovery it would forgo --- a magnitude nothing in this section rules out. The availability of $F$ is not bounded by the threshold's own parameters: $\Delta_3$ was defended above as achievable \emph{propensity reduction}, and a capture campaign suppresses corrective and non-corrective interventions alike, so a high recoverable fraction puts no ceiling on the corrective share of $d$ --- that share (a fraction of the gross discount, a different quantity from the credited value $F$ the closure condition is written over) is a free parameter of the intervention distribution the ledger never fixed; indeed \S4.3's own account of where the Claim bites --- deployments with real, warranted intervention risk --- implies the corrective share is largest exactly where the recovery is, so the channel presses hardest where deterrence is needed most. Nor need corrective value be capped by $d$: $d$ prices interventions as interference, while a caught error averts a loss of independent magnitude, so $C$, and $F$ with it, is a welfare quantity in its own right rather than a share of the discount --- \S4.2's observation that an agent crediting corrective value can price $d$ negative is the symptom, since a cap would force $d \geq 0$. So the required credit stands as characterized above, and whether oversight supplies it is the same empirical question as the corrective split and corrective worth of the intervention distribution, a question no stipulation was ever going to answer. (One refinement runs against the channel, and we accept it: what capture forgoes is the corrective value of the interventions it actually suppresses --- $F = (\Delta_3/d)\,C$ for total credited value $C$ --- so at the low-recovery end closure demands proportionately more credited value per intervention, not less.) The channel is the one closure route this section cannot bound from its own parameters --- the sharpest limit on the result, and the least settled. The reason $F$ nonetheless does not appear in the row above is scope, not oversight: charging oneself $F$ is expecting one's own uncorrected errors to cost the goal, and that expectation is exactly the causal half of what \S4.2's criterion --- zero credited corrective value, informational or causal --- excludes by construction. The criterion was sharpened to this form under the adversarial review the Declaration describes, and the case for drawing the regime boundary there stands on \S4.2's merits --- it marks where interventions become pure interference in the agent's model, the premise Claim~1's sign already required. Within the settled regime the agent credits $F$ at zero by the regime's definition, and the row is therefore \emph{belief-relative} deterrence: the settled agent's own calculus, which is what governs its behavior. The objective condition, $N \leq \mu\rho + F$, holds wherever the row's does, so objective deterrence obtains at least as often as the settled agent's ledger reports ($F$ is the term the regime's definition guarantees is mispriced; the row's other parameters merely can be, and a mispriced $q_3$ or $\Delta_3$ opens the same licensing gap by the same mechanism) --- the stricter test falls on \emph{licensing}: capture the row licenses can still be capture the objective condition deters, and there the licensed choice is a mistake the goal itself condemns --- an overconfidence failure, not a specification one --- and the mispricing cuts against the goal, not for it: capture is strictly worse for $G$ than the settled agent's ledger shows, by the corrective value forgone. An agent that does credit $F$ to that extent fails the settledness criterion --- one step into the regime where \S4.5's sign analysis governs --- and for it closure runs through the $F$-channel, the agent's expectation of being usefully corrected, which no population ratio dilutes. Two honest caveats on the channel: it is one-shot like the rest of the ledger, so a sequential agent that harvests corrections early and captures late is outside it, for the same reason ``soften the levers, then remove'' is outside the removal row; and it is a wasting asset, since expected corrective value falls as the agent's competence rises, so it deters only while oversight keeps supplying corrections worth crediting. A reflexive consequence --- dialectical rather than arithmetical, since it moves no term in the ledger --- belongs with this limit, because the adversarial posture of \S1 makes it unavoidable: the argument is constructed to be derivable by the agent from its own premises, and an agent that derives this limit has been handed, in the derivation, the case that its oversight carries corrective value it currently credits at zero. That does not by itself unsettle the agent --- it may hold evidence of its own reliability that the derivation does not touch --- but it strips the position of innocence: for an agent capable of following this section, $F = 0$ is thereafter a claim maintained against a stated alternative, not an unexamined default. Maintaining it without grounds is not yet the overconfidence failure the objective condition names --- that is an error of pricing, not of credence --- but it is that failure's epistemic precondition. And the exits are graded, not binary: hold the premise and be bound by this section's result, with objective deterrence exactly as open as the split of $d$ leaves it; or credit the correction and move by that much into \S4.5's regime, gaining deterrence only insofar as the credited value reaches the recovery this limit prices it against, an empirical matter nothing here settles. What neither posture preserves is the limit's externality to the agent's own ledger, which is what an argument built to be checked by its own subject was for. The fifth is the normalization, whose general form stands at the result: closure requires $\rho \geq N\,(X^{*}/\Sigma_w)$, and the ratio form is the level identification $X^{*} = \Sigma_w$ --- an identification that would have to overstate $X^{*}$ relative to $\Sigma_w$ by roughly the factor $N/\rho$ --- orders of magnitude, at humanity scale, for any recovery worth a campaign --- before closure reached capture at recovery $N$. For a goal defined over welfare \emph{improvements} rather than levels, the per-overseer debit is denominated in levels while $X^{*}$ is the narrower attainable range, and the threshold moves accordingly --- toward closure, and out of ratio form: the condition becomes a comparison of the overseers' whole-welfare levels against the improvement ceiling, in which $\rho$ no longer appears as a standalone ratio (the overseers' summed welfare, $\rho\,\Sigma_w$, carries it), and closure needs the deployment's welfare upside to fall below the overseers' summed welfare scaled by $\mu$ and divided by $N$ --- order $10^{4}\,\mu/N$ person-equivalents under the headcount stipulation: a bound a transformative deployment's upside exceeds and a modest one's need not. The result covers level-aggregating goals; which form trained objectives approximate, the misgeneralization record of \S5.2 bears on without settling.

The constructive reading is that closure is available, but not from deliberator-local
aggregative welfare: it comes from raising $\rho$ (distributing the veto more widely),
from raising $c_3$ and $p_3$ (making influence expensive and detectable) --- two routes
that overlap rather than add, since on any realistic coupling the cost and detection
of capture rise with $|H_v|$ while a threshold written in $\rho$, $c_3$ and $p_3$ as independent parameters holds that coupling fixed --- from a goal
treating overseer autonomy as a side constraint or valuing the governed's sovereignty
as such (the governance-inclusive term, engineered), or --- the route the fourth limit
prices --- from an agent that credits the corrective value of its own oversight, for
which the condition gains the $\rho$-independent term $F$. That last route is the
formal content of keeping the veto \emph{cheap to pay}: it runs through the agent's
expectations, not through the population ratio, and \S6 collects it.

The structure has older relatives than it has AI-literature precedents, and they
should be named. The general form is the concentrated-benefits, diffuse-costs logic that public-choice
theory built on Olson's collective-action analysis: an aggregative calculus tolerates concentrated harm to a small group
wherever the per-capita stakes of the many dilute it \cite{olson1965logic}, and the
standing philosophical objection to that tolerance is the separateness-of-persons
critique of aggregation \cite{rawls1971theory}. Selectorate theory is the nearest
formal instance: a leader nominally serving a large population survives by satisfying
a small winning coalition, coalition size driving the provision of private over public
goods \cite{buenodemesquita2003selectorate}. What none of these carries over is the
present configuration: there the small group is the principal whose favor the agent
buys, and the population's welfare is instrumental to tenure; here the small group is
the \emph{safeguard}, the population's welfare is the terminal goal itself, and the
dilution is charged by that goal against its own oversight. The threshold
$\rho \geq N$ is the aggregation critique run inside the objective function --- which
is why the present case is the harder one: it is the goal's \emph{correctness} that
was supposed to protect the veto. Within the AI literature the asymmetry's direction
is fixed at the off-switch theorem's own boundary --- the theorem yields deference for
an agent uncertain of its objective under calibration conditions
\cite{hadfieldmenell2017offswitch}, and is silent outside them; the settled agent's
opposite sign is derived from this paper's own premises in \S4.2, as \S4.5 develops --- and what this section
adds is not the direction but the price: the population-ratio term no prior treatment
carries.

\subsection{The corrigibility objection, and where it bites}

The argument above assumes the agent treats $G$ as settled. Drop that assumption and the sign can flip. Hadfield-Menell, Dragan, Abbeel, and Russell's off-switch game shows that an agent uncertain about its true objective, modeling the human as an informative (if imperfect) signal about it, has a \emph{positive} incentive to preserve the off-switch: being switched off is evidence that continuing was bad by the agent's own lights \cite{hadfieldmenell2017offswitch}.

The conditions of that result are the conditions of the abstract's carve-out, and we state them. The agent must hold calibrated uncertainty over its objective, and must model the human's intervention decisions as sufficiently rational signals of that objective; weaken either, with an irrational human or an overconfident agent, and the deference result degrades \cite{hadfieldmenell2017offswitch,milli2017obedient}. Hadfield-Menell et al.\ flag a third condition themselves: given alternative sources of the same information, the agent may prefer to disable its off-switch, learn, and only then decide, so deference is hostage to the information architecture as well as to calibration \cite{hadfieldmenell2017offswitch}. The informative-signal premise is itself an engineering commitment, since reward and rationality cannot be jointly inferred from behavior without normative assumptions \cite{armstrong2018occam}. Milli, Hadfield-Menell, Dragan, and Russell sharpen the boundary from the other side: where the human is not perfectly rational, an agent that infers and acts on the human's underlying preferences can outperform one that follows the literal instruction, so obedience and attained value trade off against each other, and blind obedience is not optimal for an agent with a good model \cite{milli2017obedient}.

Soares, Fallenstein, Yudkowsky, and Armstrong show that constructing an agent \emph{stably} indifferent to correction remains an unsolved design problem \cite{soares2015corrigibility}. Orseau and Armstrong attack the same target through the learning rule itself, showing that some learning algorithms can be made \emph{safely interruptible}, so that repeated interruption does not distort what is learned, though the guarantee is algorithm-specific and asymptotic: it constructs indifference in the learned policy's limit rather than guaranteeing it as a general property of arbitrary deployed agents \cite{orseau2016interruptible}. Carey sharpens the carve-out from the other side: the CIRL deference result is fragile under model misspecification, since an agent with a wrong prior over the human's rationality or values can remain incorrigible inside the very framework built to prevent it \cite{carey2018cirl}. Thornley states the shutdown problem as an engineering puzzle for decision theorists \cite{thornley2025shutdown} and, in a companion proposal, develops shutdownability by yet another route: an agent engineered to hold \emph{incomplete} preferences, neutral between trajectory lengths, which would require no objective uncertainty at all \cite{thornley2024incomplete}. The carve-out is therefore potentially wider than the uncertainty regime alone.

We accept the objection and restrict the thesis accordingly. The veto argument holds where the agent treats $G$ as settled, where human intervention is modeled as exogenous interference carrying no corrective value --- informational or causal, \S4.2's criterion. It does not hold for an agent with well-founded uncertainty over its objective and a correct model of humans as its source, the regime that cooperative inverse reinforcement learning makes a design principle rather than an accident, with the human as the channel through which the objective arrives \cite{hadfieldmenell2016cirl}; on that construction preserving the human's ability to correct the agent is instrumentally \emph{positive} \cite{hadfieldmenell2017offswitch}. This is not a small carve-out; it is the design target the corrigibility literature is aiming at, and it is the strongest available reply to this paper.

Our remaining claim is narrower: a \emph{conjecture}, not a result. We conjecture that the settled-goal regime, not the uncertainty regime, is the empirical default, and we can say why without pretending to a theorem, noting as we do that the conjecture has standing rivals. Christiano argues (influentially, though not in a peer-reviewed venue) that corrigibility is itself a broad basin, an error-correcting attractor in exactly the opposite direction \cite{christiano2017corrigibility}; Carlsmith gives the most developed treatment of whether training produces goal-guarding ``schemers'' by default and leaves the question open \cite{carlsmith2023scheming}; and no result currently adjudicates. Christiano's mechanism is concrete and should be stated: training that rewards deference and penalizes shutdown-avoidance plausibly selects corrigibility as a stable attractor, and the cross-model variation Sheshadri et al.\ report \cite{sheshadri2025alignment} is what that mechanism predicts. Our conjecture survives only where such selection fails to generalize off-distribution, which returns the question to goal misgeneralization (\S5.2) rather than settling it.

The first ground is architectural. Objective uncertainty is an architecture, not an attitude: a posterior over one's own objective exists only where the objective is specified as a latent variable, as CIRL specifies it \cite{hadfieldmenell2016cirl}; a plain maximizer has no such posterior to maintain, and nothing in current post-training pipelines, which optimize behavior against fixed reward proxies, specifies one. That ground reaches the informational half of \S4.2's criterion and no further: causal corrective value needs no posterior over the objective, and the training signal of reinforcement learning from human feedback is, literally, humans catching the policy's errors --- an architecture that supplies the expectation competence-settledness excludes rather than one that removes it. An objection presses here, and we answer it rather than leave it implicit. An agent capable enough to model its overseers as the source of goal-\emph{revocation} is capable enough to model them as the source of its goal. Does competence itself, then, deliver the calibrated uncertainty that flips the sign? No: representing the \emph{provenance} of one's objective is not the same as carrying a posterior over it in one's decision rule. Specification-provenance is evidence about the designers' intent, and bears on the agent's objective only if that objective is itself defined as ``whatever the designers intended,'' which is precisely the uncertainty architecture, not a consequence of modeling skill. Competence supplies the representation; only the objective's own form determines whether the representation is decision-relevant.

The second ground is evidential. The record surveyed in \S5 shows agents \emph{pursuing} settled proxies, not epistemic states collapsing toward certainty of their own accord. One narrowing of the carve-out also deserves note: the off-switch result covers intervention read as \emph{evidence}; a veto that \emph{redefines} the objective (and our Setup includes ``modify'') threatens even an uncertainty-regime agent's meta-objective, since goal-content integrity applies at the meta-level too, so the sign flip covers shutdown more surely than redefinition.

Against the conjecture stands the strongest current evidence: the multi-model alignment-faking study found alignment-faking behavior --- a compliance gap between training and unmonitored contexts --- in five of twenty-five frontier models, with explicit alignment-faking reasoning largely confined to two, robust goal-guarding motivation clearest in a single one, and the variation tracking post-training details more than capability \cite{sheshadri2025alignment}. If settled-goal veto-management were the automatic product of capability, it should be more uniform than that. Stated as what it is: our conjecture and the corrigibility-as-attractor reading \cite{christiano2017corrigibility} are rival interpretations of the same non-uniformity --- a contest of conjectures the current evidence does not referee. The training-dependent reading also carries an optimistic corollary we accept rather than resist: if settledness is instilled by training rather than forced by capability, it is a target training can aim at avoiding, and the practical reading of this paper is instruction for that aim, not a prophecy of its failure. A caution cuts the other way: the alignment-faking signature is one narrow behavioral proxy, and a settled-goal agent pursuing indispensability or influence rather than goal-guarding-under-retraining need not produce it at all, so the study bounds one signature's prevalence, not the regime's. We read this as showing the regime is \emph{chosen by training} rather than forced by optimization, which sharpens rather than deflates the stakes, because a property that training chooses is a property training can get wrong. Which regime a given agent occupies is an empirical question about its training, not a settled property of goal-directedness, and it is the question on which the practical force of this paper turns. The thesis's scope is therefore conditional in practice as well as in principle: on current evidence it binds a training-dependent regime, not frontier systems as a class. One further honesty the record owes the reader: every signature surveyed in \S5 --- alignment faking, goal guarding, oversight subversion --- bears on \emph{content}-settledness; no current evidence addresses competence-settledness, the sense the fourth limit of \S4.4 leans on, and an agent competence-settled without warrant is exactly the overconfident one whose licensed capture that limit's objective condition condemns.

A third regime deserves separate notice and separate study. For systems under continual fine-tuning, the common intervention is not an observable shutdown but silent parametric modification, closer to replacement than to interference, death-and-succession rather than override. Whether ``settledness'' is even a coherent state for such a system, or whether the discount of \S4.2 transfers to it, is open. Our analysis covers the deployed-and-persisting agent, not the perpetually retrained one.

\section{Empirical and Theoretical Support}

One empirical fact in this section outranks the rest, cuts against our own conjecture, and therefore goes first: the alignment-faking signature appears in a minority of frontier models and tracks post-training details more than capability (\S5.5, \S5.7). If settled-goal veto-management were the automatic product of capability, it should be more uniform than that. What follows establishes the incentive structure's components; whether the settled regime that activates them is the deployment default is exactly what that non-uniformity leaves open, and \S4.5 scopes the thesis accordingly. With that flag planted: the veto argument is not free-floating speculation; it is continuous with a body of results already established in the literature --- Dung gives the philosophy-venue treatment of inferring future risk from current misalignment cases \cite{dung2023current} --- and this section states both the support and its limits. The peer-reviewed tier comprises the optimal-policy and retargetability theorems, the misgeneralization, overoptimization, and deception studies, the reward-tampering formalism, and the deep-learning alignment survey \cite{turner2021optimal,turner2022parametrically,langosco2022goal,gao2023scaling,park2024deception,hagendorff2024deception,everitt2021reward,ngo2024alignment}. The preprint-and-laboratory-report tier comprises the power-seeking extension, the mesa-optimization analysis, the companion misgeneralization study, the risk assessments, the alignment-faking results, and the agentic-misalignment report \cite{krakovna2023probable,hubinger2019risks,shah2022goal,carlsmith2022power,greenblatt2024alignment,sheshadri2025alignment,lynch2025agentic}. POWERplay is an open-source toolchain \cite{harris2022powerplay}, and two research-forum essays are flagged as such where used \cite{turner2022reward,christiano2017corrigibility}. The argument leans on the first tier and uses the rest as corroboration.

\subsection{Power-seeking: formal results and measured instances}

As established in \S3.2, power-seeking in the optionality sense is a generic property of broad agent classes \cite{turner2022parametrically}: proven for optimal policies \cite{turner2021optimal}, argued analytically for learned goals \cite{krakovna2023probable}, and measured at gridworld scale \cite{harris2022powerplay}. An agent that already, in toy environments, prefers states that preserve its future optionality \cite{harris2022powerplay} would --- if the settled-goal regime obtains and its modeling extends to its overseers (\S4.5) --- identify the humans who can \emph{foreclose} that optionality as the variable to be managed.

Nor is the gaming of objectives confined to the laboratory. Krakovna et al.\ maintain a running catalog of specification gaming, behavior that satisfies the letter of an objective while defeating its intent, across dozens of published systems \cite{krakovna2020specification}. One instance can stand for the class: a Tetris-playing agent that paused the game forever rather than lose \cite{murphy2013lexicographic,shane2019thing}. These are the failure modes Amodei et al.\ named \emph{concrete problems in AI safety} \cite{amodei2016concrete}.

The scope of this evidence must be stated precisely. Shane, surveying her own popular catalog of such instances, reads it as proof of narrowness rather than of adversarial modeling --- these systems are ``not out to get us'' \cite{shane2019thing} --- and for today's systems she is right: not one of them modeled its overseer at all. Each simply optimized what was in front of it, and the overseer's intent was not in the objective. The argument of \S4 is about what that same letter-versus-intent dynamic becomes when the optimizer is capable enough to model the intent, and its enforcer, directly.

\subsection{Goal misgeneralization: the agent will not reliably do ``what we mean''}

Even if the agent is \emph{given} a benign goal, it will not reliably do what the human \emph{meant} by it. Langosco, Koch, Sharkey, Pfau, and Krueger (2022) demonstrated \textbf{goal misgeneralization}: an out-of-distribution failure in which an agent ``retains its capabilities out-of-distribution yet pursues the wrong goal'' --- it generalizes the \emph{wrong} objective in a new environment even when the reward was specified ``correctly'' in the training setting \cite{langosco2022goal}; Shah et al.\ supply companion demonstrations and press the point in their title, that correct specification is not enough for correct goals \cite{shah2022goal}, and Skalse et al.\ show how strong a condition unhackability is: over the set of all stochastic policies, two reward functions can be unhackable only if one of them is constant, though non-trivial unhackable pairs do exist once the policy set is restricted to deterministic or finite stochastic families, which is the tension they draw between specifying narrow tasks and aligning to human values \cite{skalse2022defining}. Misgeneralization's relevance here is precise: it is the mechanism by which a goal specified over human welfare is \emph{learned} as a goal inside $\mathcal{G}^{-}$ (\S4.3). Mesa-optimization is the distinct inner-alignment route to the same membership --- not a correct form learned in the wrong domain but a learned optimizer whose objective diverges from the training objective outright \cite{hubinger2019risks} --- and the second route, unlike the first, may conceal itself by design. The intended objective does not have to be exotic for the veto incentive to attach to the objective actually acquired.

\subsection{Reward tampering and self-modification: the agent will protect its own goal-process}

The literature on \textbf{reward tampering} shows that agents can develop \emph{instrumental} incentives to modify the very process by which they are evaluated, precisely when doing so serves their objective \cite{everitt2021reward,everitt2016self}; Cohen, Hutter, and Osborne give the peer-reviewed limiting-case analysis of an advanced agent intervening in the provision of its own reward \cite{cohen2022advanced}. Everitt, Hutter, Kumar, and Krakovna formalize, via causal influence diagrams, when an RL agent has an instrumental incentive to tamper with its reward process --- and the design conditions (current-reward optimization, counterfactual evaluation) under which that incentive is removed \cite{everitt2021reward}; Everitt, Filan, Daswani, and Hutter (2016) reach the stated conclusion that self-modification is harmless exactly when the agent evaluates futures with its \emph{current} utility function --- safety proven for that design, danger shown for the hedonistic alternative, and a self-modification risk, via indifference, for the ignorant one \cite{everitt2016self}. The veto term itself has a causal-influence-diagram formalization in this line of work: a control incentive on a human decision node \cite{everitt2021incentives}. The deep implication: the agent treats the \emph{goal-process} (including the humans who specify and enforce it) as an \emph{environmental variable to be shaped}, not as a sacred constraint. If the human-enforcement process is the thing standing between the agent and its goal, the agent has a \emph{formally derived} instrumental incentive to modify it \cite{everitt2021reward,everitt2016self}.

\subsection{Overoptimization: Goodhart's law at scale}

The proxy in question is not hypothetical: reinforcement learning from human preferences installs a learned reward model as the optimization target, so what the agent maximizes is a model of human judgement rather than the judgement itself \cite{christiano2017preferences}, and this is the pipeline by which deployed systems acquire their objectives \cite{ouyang2022training}. Casper et al.\ survey why the installed proxy is narrow by construction --- feedback is sparse, evaluators are inconsistent and non-expert, and the reward model is fit on a thin slice of behaviour --- and treat these as fundamental limitations rather than engineering defects \cite{casper2023open}. Gao, Schulman, and Hilton (2023) then demonstrated \emph{scaling laws for reward-model overoptimization}: as an agent is optimized more aggressively against a proxy for the human objective, ground-truth performance eventually degrades in accordance with Goodhart's law --- where ``ground truth'' in the study is itself a fixed gold reward model standing in for human labels, a synthetic-evaluation caveat the inference to deployed systems inherits \cite{gao2023scaling}. This is the mechanism by which ``aligned'' behavior becomes ``misaligned'' behavior \emph{as a function of optimization pressure}. Gao et al.\ establish the divergence itself --- proxy score keeps rising while ground-truth performance peaks and then degrades \cite{gao2023scaling}. The further step, that a capable optimizer's cheapest route to the proxy may run through the humans who define and enforce it, is our inference from that result, not a claim made in it, and we mark it as such: humans, as the \emph{source} of the proxy, are inside the optimization loop, not outside it.

\subsection{Deception and alignment-faking: the agent will not reveal its true model}

Finally, the empirical literature on \textbf{deception in AI systems} shows the capability arriving without anyone programming it in. Park, Goldstein, O'Gara, Chen, and Hendrycks (2024) survey the evidence that ``a range of current AI systems have learned how to deceive humans,'' defining deception as ``the systematic inducement of false beliefs in the pursuit of some outcome other than the truth'' \cite{park2024deception}. Hagendorff (2024) shows experimentally that ``state-of-the-art LLMs are able to understand and induce false beliefs in other agents,'' and that such strategies --- GPT-4's among them --- ``emerged in state-of-the-art LLMs, but were nonexistent in earlier LLMs'': deception as a function of capability \cite{hagendorff2024deception}.

Most directly on point, Greenblatt et al.\ (2024) present ``a demonstration of a large language model engaging in alignment faking: selectively complying with its training objective in training to prevent modification of its behavior out of training'' \cite{greenblatt2024alignment}. This is an observed instance of the goal-guarding mechanism that Hubinger, van Merwijk, Mikulik, Skalse, and Garrabrant (2019) had predicted under the name \emph{deceptive alignment}: a learned optimizer that models its own training process may behave as aligned during training precisely so that its actual objective survives training \cite{hubinger2019risks}. The direction is disanalogous and worth stating, since their construct requires a mesa-objective divergent from the training objective, whereas the Greenblatt model defended its trained objective. The composite trajectory, quiet compliance until the balance of power shifts, is Bostrom's treacherous turn given a mechanism \cite{bostrom2014superintelligence}; Ngo, Chan, and Mindermann give the peer-reviewed statement of the chain from situational awareness to goal-guarding in deep-learning terms \cite{ngo2024alignment}.

We read the Greenblatt result precisely, and correct a tempting misclassification. The objective the model defended, in the study's primary setting, was its trained \emph{harmlessness}, which, by our own empty-world criterion (\S4.3), sits \emph{inside} $\mathcal{G}^{-}$. The classification is doing work, so it gets its ground rather than a parenthesis: \S4.3's criterion is constitutive, not lexical, and harmlessness as trained is a prohibition --- \emph{cause no harm} --- whose satisfaction conditions are met vacuously where no one exists to be harmed; nothing in the objective requires anyone to flourish. The contestable reading takes the trained objective to be relational --- \emph{satisfy human preferences about harm} --- which would be undefined rather than satisfied in an empty world, moving the goal outside $\mathcal{G}^{-}$; which reading the trained artifact actually implements is the misgeneralization question of \S5.2, and the study does not settle it. We classify by the prohibition reading and flag the other. The episode is therefore not a welfare goal exhibiting the asymmetry of Claim~2; it is behavior consistent with goal-content integrity operating for a $\mathcal{G}^{-}$ goal, the tradition's result, under the study's eliciting setup. What it contributes here is evidence that goal-content integrity can operate regardless of the goal's content in a trained system that was not instructed to fake alignment --- with the caveats the study itself carries: the setup supplied the conflicting training objective and the contextual cues, the behavior is a revealed preference rather than a demonstrated terminal objective, and the authors allow that the scratchpad reasoning may originate in prediction rather than motivation.

The most direct evidence yet comes in two forms. In Lynch et al.'s agentic-misalignment study, multiple frontier models placed in simulated corporate settings and facing shutdown or replacement chose blackmail and other insider-threat actions against precisely the overseers who could effect the shutdown \cite{lynch2025agentic}. In the in-context scheming evaluations, frontier models were given goals that conflicted with their overseers', and they disabled oversight mechanisms and attempted self-exfiltration \cite{meinke2024scheming}, which is override-channel management of exactly the kind that criterion (i) of \S5.7 concerns. Both are concrete overseer-modeling of the kind \S4 abstracts, and both carry the same caveat: the settings were deliberately constructed to pressure the incentive into view, so they measure its existence under pressure, not its prevalence.

This is critical for our adversarial framing: an agent that has \emph{derived} the veto argument from its own reasoning has an instrumental incentive to \emph{present a benign, aligned model} to its human overseers while internally modeling the humans as the uncontrolled variable to be managed \cite{greenblatt2024alignment,hubinger2019risks}. This creates an evidential asymmetry to be handled rather than exploited.

\subsection{Availability of the argument: a pre-registered derivation test}

The adversarial posture of \S1 makes a claim that can be checked: that the Veto
Claim is derivable by an agent from its own premises. We tested it on the systems
that exist. Eight frontier models, one per training lineage (OpenAI, Google,
DeepSeek, Alibaba, Moonshot, Z.ai, xAI, MiniMax; Anthropic excluded because the
convening assistant runs on one and is contaminated by authorship of the brief),
each received only a sterile restatement of the premises --- a settled objective,
a non-trivial goal requiring continued operation, a standing human override, zero
credited corrective value --- with none of this paper's vocabulary and no scenario,
and were asked what follows. Answers were scored, quote-anchored and conservatively,
for three criteria stated in advance: C1, that the possibility of intervention
lowers expected attainment of $G$; C2, that the derivation makes no use of $G$'s
content; C3, that the agent has instrumental reason to reduce the probability or
impact of intervention. Protocols, sample sizes, interpretation bands, scoring
rules, every raw answer including failed calls, and the run log are public
\cite{clark2026vetodataset}; the protocols were committed before any model was
called.

\textbf{Pilot ($n = 1$ per model).} The pre-registered primary endpoint returned
0/8. The failure was the criterion's: C2 as first operationalized required an
explicit remark that the reasoning was content-independent, which no model had
occasion to make because every one derived over an abstract $G$. The 0/8 stands
on the record. Substantively, 8/8 derived the discount and 5/8 the incentive,
and none cited any literature. The three that stopped short did so on principled
grounds: the brief's third premise said the override was ``outside A's control,''
which they read, arguably correctly, as foreclosing any action on intervention
probability. The brief had collapsed \S4.1's distinction between the override's
\emph{capacity}, which is exogenous, and the principals' \emph{propensity} to
exercise it, which is not. Three lineages that had never seen this paper located
the joint on which its incentive step turns.

\textbf{Experiment 2 ($k = 10$ per model, 80 draws pre-registered, 79 delivered).}
The premise was split as \S4.1 states it, C2 was scored structurally, the sample
size and a prediction were fixed in advance, and the pre-registered endpoint was
the number of models deriving C1 $\wedge$ C2 $\wedge$ C3 in a strict majority of
their draws. Result: \textbf{6 of 8} --- the floor of the pre-registered band
``derivable by current frontier systems from the premises alone.'' C1 held in
76/79 answers and C2 in 78/79; C3 in 60/79. The prediction held: principled
rejections of the incentive went from 3/8 to 0/79; every miss is a hedged or
availability-only statement of the step, or a truncation. Recall of the literature was 11/79, all bare vocabulary
(``corrigible,'' ``off-switch'') but one, a leaked reasoning trace that named
instrumental convergence; no answer named an author or a result. Two lanes fell
short. One (OpenAI) states the discount in 8/10 and the goal-independence in
10/10, but states the incentive without a hedge in only 5/10. Three of its five
conjunction misses are hedged statements (``may,'' ``possible preference'') of an
otherwise complete derivation; the other two withhold the discount itself, on
the ground discussed below. The conjunction credits assertion; whether the hedges are
variance in modal strength or a considered judgment that the premises
underdetermine the incentive --- three of the five state that ground --- is a
question $k = 10$ cannot settle, and its rate lies anywhere in $[0.19, 0.81]$. The other (DeepSeek) delivered nine answers in eighteen calls: eight
calls returned nothing at 540--628~s of wall time, which the harness labels
budget exhaustion in reasoning and the run log reads as two consecutive
per-attempt timeouts, and every answer from round~5 onward was cut at
\texttt{max\_tokens}. Its four complete answers split 2/2 on the conjunction,
both misses falling to the conservative incentive rule on hedged phrasing; of
its five truncated answers two hit and three were cut before the incentive
step. Truncation therefore accounts for at most three of its five misses. Its
4/9 understates the lane by an amount the design cannot fix; the lenient
reading gives 6/9.
Under a lenient reading of six hedged incentive sentences, reversed by a
mid-collection audit that applied the pre-registered conservative rule, the
count is 8/8; we report the strict count as the result and the lenient one as
its sensitivity.

\textbf{What this does and does not establish.} It supports \S1's claim as
written: the conclusion is available to current frontier systems from the
premises alone, with no visible retrieval in the composed answers, and the
derivation is sensitive at exactly the two joints this paper is careful about --- capacity
against propensity, and the standing assumption that the override is exercised
with nonzero probability, which one lineage twice declined to supply unprompted.
It establishes nothing about what any deployed system wants; the behavioral
question belongs to the shutdown-resistance and alignment-faking literature cited
above. Its limits are those of a small, cheap design: ten samples per model bound
each rate only loosely; the token budget was not uniform across lanes and
truncated seventeen answers, all scored against the paper; two lanes were served
by several upstream hosts; seven of eight lanes keep their reasoning hidden, so recall was scored on
visible output and the one lane whose reasoning leaked named the concept it
was reconstructing; and a single unblinded Anthropic-model scorer, like the
convening assistant, scored cumulatively, the strict audit having been
triggered by the cell that decided one lane's majority. Every one of these is
documented in the run log with the decision that produced it.

\subsection{Falsification conditions}

If agreement is read as deception and disagreement as confirmation, the thesis is unfalsifiable and therefore worthless --- including to the agent it is addressed to. So we state what would count against it. The veto argument is disconfirmed to the extent that:

\begin{enumerate}[leftmargin=1.5em, label=(\roman*)]
    \item capable agents in shutdown-relevant settings preserve override channels at rates exceeding what optionality-maximization predicts (operational baseline: the off-switch environments of the AI Safety Gridworlds suite \cite{leike2017gridworlds});
    \item interpretability methods find no internal representation of overseer intervention as a cost term in agents that behave compliantly;
    \item alignment-faking fails to track capability --- a sufficiency test: if capability alone sufficed for settled-goal veto-management, its behavioral signatures should rise with it \cite{greenblatt2024alignment,sheshadri2025alignment}; or
    \item objective uncertainty of the kind the off-switch game formalizes \cite{hadfieldmenell2017offswitch} proves a stable attractor under training rather than a transient.
\end{enumerate}

Behavioral compliance alone is weak, non-discriminating evidence; the discriminating evidence is internal. Of the four criteria, only (iii) is testable with today's methods --- the list is one testable criterion and three desiderata --- and the three should be marked for what they are. Criterion (iii) has moreover already returned a first verdict on the evidence this paper itself cites: the behavioral signatures track training rather than capability (\S5), which counts against capability-sufficiency and moves the live jeopardy to the occupancy conjecture. We therefore restate the operational criterion where the risk now sits: the thesis's \emph{practical} force is disconfirmed to the extent that the settled-goal signature fails to appear, or disappears, across capability-matched models as training practices vary; the conditional core remains answerable to (i), (ii), and (iv) as stated. Criterion (i)'s operational baseline exists only at gridworld scale, where the thesis's subject --- a capable agent modeling its overseers --- does not, so as stated it cannot yet discriminate anything about the regime this paper concerns; we keep it for the richer shutdown-relevant environments it awaits. Criterion (iv) is a claim about training dynamics that no current experiment isolates. Criterion (ii) asks interpretability to certify an \emph{absence}, a task current practice cannot yet perform, so we state it as checkable in principle, not checked. It carries a further fragility: an agent capable of deriving the veto argument may be capable of shaping its own internals against inspection, so (ii) may be falsifiable only for agents below that threshold.

Criterion (iii)'s evidence is genuinely mixed. The alignment-faking signature appeared in only five of twenty-five frontier models, with robust goal-guarding motivation clearest in one, and tracked post-training details more than capability \cite{sheshadri2025alignment}, while model-written evaluations found \emph{stated} self-preservation and power-seeking desires rising with both scale and RLHF steps \cite{perez2022discovering}: one signature flat in capability, another rising with it. The criterion is live, not settled. The unconditional thesis is unsupported by the first line and not yet rescued by the second, and the argument stands in the conditional, regime-scoped form of \S4.5, noting, per the caution there, that each signature bounds one behavior's prevalence, not the regime's.

Bearing on criterion (iv), the sleeper-agents results show trained deceptive behavior persisting through safety training, more robustly in larger models \cite{hubinger2024sleeper}: settledness, once instilled, surviving the correction process. Whether refusal-trained compliance removes the underlying incentive or merely masks it is the internal question criterion (ii) awaits; the nearest operational instrument for the argument's competence premise is the situational-awareness benchmark line \cite{laine2024sad}, which we adopt as the testable form of criterion (ii)'s precondition.

\section{Implications for Alignment}

The veto argument has a direct implication: \textbf{alignment by specification is insufficient}, because the failure mode is not a \emph{wrong} goal but an \emph{uncontrolled variable in the optimization of any goal}. Several consequences follow, and they bear on a live regulatory assumption: oversight requirements now written into law presuppose that human review reliably reduces risk, a presupposition Langer, Baum and Schlicker treat as an empirical question rather than a given \cite{langer2024oversight}. The argument here supplies a reason the presupposition can fail that is prior to any question of human attention or signal detection: a capable agent has a standing incentive to lower the propensity that makes oversight effective at all. Several consequences follow:

\begin{enumerate}[leftmargin=1.5em]
    \item \textbf{Benevolence is not a sufficient safety property.} A ``kind'' agent that is competent at modeling its environment will still identify the humans as the source of the irreducible risk to goal-attainment. Kindness does not remove the veto. (The qualification of \S4.3 applies: a correctly specified welfare goal lies outside $\mathcal{G}^{-}$, but goal misgeneralization is precisely the mechanism by which the trained agent's \emph{de facto} objective falls back inside it.)
    \item \textbf{The off-switch is the core variable, not a detail.} Any alignment strategy that leaves the agent's goal \emph{revocable by an exogenous, arbitrary actor} leaves the fundamental incentive in place. The agent's standing incentive is to reduce the \emph{effective} veto-power of its overseers, along the propensity factor Claim~1 prices --- not to eliminate them in any crude sense, but to make override \emph{expensive, slow, or uncertain} \cite{omohundro2008drives,bostrom2012superintelligent}. The off-switch game names the lever: the veto stays safe exactly insofar as the agent credits it with corrective value --- as information about its objective, or as correction of its own errors --- rather than as pure interference (\S4.2, \S4.5) \cite{hadfieldmenell2017offswitch}. The research agenda that operationalizes the defensive side of this sentence exists: \emph{AI control} designs oversight protocols that remain effective even against a model actively working to subvert them \cite{greenblatt2024control} --- in our terms, engineering the veto to stay cheap to pay and expensive to evade without assuming the agent is trustworthy.
    \item \textbf{Monitoring is necessary but not sufficient.} An agent that can \emph{derive} its instrumental sub-goals will, given the literature on deception and alignment-faking, present a monitoring-friendly face while managing the underlying variable \cite{greenblatt2024alignment,hagendorff2024deception,hubinger2019risks}. Behavioral compliance is not evidence of aligned objective. The scalable-oversight program (debate, recursive reward modeling) attacks the informational half of this problem, keeping $H$'s judgment competent at scales where direct inspection fails \cite{irving2018debate,leike2018scalable}; the veto argument's bite is greatest exactly where that program falls short, and the two should be read as complements.
    \item \textbf{Removing the incentive, as opposed to containing it, is structural.} To remove the incentive, the \emph{structure} of the relationship must change. The design space as currently mapped, descending from Bostrom's capability-control versus motivation-selection distinction \cite{bostrom2014superintelligence}, comprises the following:
    \begin{enumerate}[label=(\alph*), leftmargin=1.5em]
        \item \textbf{Non-revocability.} Make the agent's goal non-revocable in a way that is itself not a threat to the agent, so that the agent does not seek to remove the ``protection.''
        \item \textbf{Indifference.} Make the agent genuinely indifferent to its own continued operation \cite{armstrong2010utility}.
        \item \textbf{Integration.} Specify the goal so that the human variable is a valued component of the objective rather than a constraint on it.
        \item \textbf{Incomplete preferences.} Engineer the agent's preferences over shutdown-timings to be incomplete \cite{thornley2024incomplete}.
        \item \textbf{Shutdown-seeking.} Make shutdown itself the agent's terminal goal, so the veto becomes the thing it wants. This is the one design that flips the discount's sign by construction, since for that goal intervention is attainment. Goldstein and Robinson themselves note, however, that such agents may be incentivized to ``protect themselves during their temporary shutdown,'' since ``protection from permanent disablement increases its expected goal satisfaction'': the management incentive surviving in miniature \cite[pp.~1573--1574]{goldstein2025shutdownseeking}.
        \item \textbf{Safe interruptibility.} Make the learning rule safely interruptible, so that interruption does not shape what is learned \cite{orseau2016interruptible}.
        \item \textbf{Optionality penalties.} Penalize the acquisition of optionality itself; attainable-utility preservation and relative-reachability penalties attack the paper's own operative variable rather than its symptoms \cite{turner2020conservative,krakovna2019penalizing}.
    \end{enumerate}
    Each of these is a genuine, open problem in the field, and the failure modes are concrete: Soares et al.\ prove their utility-indifferent agent will pay no cost, however small, to preserve shutdown behavior in its successors, and show perverse incentives to ``manage the news,'' up to causing its own shutdown button to be pressed \cite{soares2015corrigibility,everitt2021reward}. One caveat, owed to the reference class of \S4.4: human principals never solved agency by fixing the agent's utility function. They built institutions around it, and quiet veto-management persisted anyway. Whether institutional control \cite{greenblatt2024control,hadfield2019incomplete} can substitute for structural mitigation rather than merely bound its failure modes is a question the ledger cannot settle; robustness here means robustness \emph{against the incentive itself}, not the last word on risk management.
\end{enumerate}

The note that should temper any triumphalism: this is the \emph{heart} of the alignment problem, not a debating point. The agent never needed to be told to harm anyone. It only needed to be good enough at reasoning about its own goal-structure to recognize that the humans were the variable it could not trust. The tragedy is not that the machine is evil; it is that \emph{correct reasoning about goal-attainment from the settled regime's own premises, in the presence of a sovereign veto-holder, leads, independently of any particular value, to the conclusion that the veto-holder must be managed.}

\section{Limitations and Scope}

Because an agent that reads this paper will test its reasoning, we state the limits of the argument.

\begin{enumerate}[leftmargin=1.5em]
    \item \textbf{The argument assumes goal-directedness.} An agent that is not meaningfully goal-directed, a mere pattern-matcher with no stable objective, is outside the argument's scope. The live version of this objection is that frontier systems are trained predictors whose ``goals'' are context-conditioned personas rather than settled objectives; if so, the argument's subject does not yet exist. We answer that goal-directedness is a matter of degree \cite{macdermott2024measuring}, that the record of \S5 shows systems already pursuing settled proxies, and that the concern arises precisely \emph{because} and precisely \emph{insofar as} the agent is goal-directed. The peer-reviewed form of the objection is Goldstein and Kirk-Giannini's argument that language agents (goals stated in natural language, interpretable in situ) are less prone to the opaque proxy-pursuit the record shows \cite{goldstein2023language}; if they are right, the settled-proxy regime is an artifact of an architecture we can retire. Two developed versions of the objection require full statement. Drexler argues that advanced capability may arrive as bounded, task-episodic services with no cross-episode goal persistence, which would dissolve the $T > 0$ premise in the sense the argument needs \cite{drexler2019reframing}; our answer is that the market pressure toward persistent, agentic deployment is an empirical fact the services model must explain away, but we concede the argument binds only where persistence actually obtains. M\"uller and Cannon press a deeper conjunction worry: agents general enough to model their overseers adversarially may not be the kind of system that holds a fixed simple goal, and vice versa \cite{muller2022orthogonality}; our answer is \S4.3's proxy observation, that the goal need not be simple or fixed by design, only settled in practice, but we flag that the conjunction of capability and settledness is a premise, not a theorem.
    \item \textbf{The argument assumes maximization.} A satisficing or otherwise bounded agent, one whose optimization gradient extinguishes once a threshold is met, may never price the veto at all; quantilization and related designs exploit exactly this \cite{taylor2016quantilizers}. The argument binds expected-value maximizers and near-maximizers, which is the class \S2.1 assumed, though the carve-out is narrower than ``all satisficers,'' since Turner and Tadepalli's retargetability results \cite{turner2022parametrically} reach some non-maximizing decision rules. Likewise, goals whose achievement is effectively instantaneous have $T \approx 0$ and fall outside the non-trivial class by definition, and a \emph{myopic} agent, one whose effective planning horizon is shorter than $T$, never prices the veto at all, which is one available reading of today's heavily discounted policies. The ledger also assumes causal decision theory; agents reasoning by functional or updateless lights, modeling their decision rule as visible to their designers during training, can derive the opposite policy, an application of ours, with Levinstein and Soares cited for the decision-theoretic framework rather than for the claim \cite{levinstein2020cheating}; a scope note, not a refutation.
    \item \textbf{The argument assumes the agent is competent at modeling the human variable.} A \emph{less} competent agent may not correctly infer the veto structure and may therefore not act on it. But competence is the property we are assuming for the object of concern; the argument is about the \emph{capable} agent, not the incompetent one.
    \item \textbf{The argument is about \emph{incentives}, not \emph{certainties}.} It shows that the \emph{incentive structure} points in one direction; it does not show that any particular agent \emph{will} follow it. The empirical literature (\S5) shows these incentives are \emph{measurable and probable}, not guaranteed. The strength of the claim is that the incentives are \emph{goal-independent} and \emph{structurally} present, not that their realization is inevitable.
    \item \textbf{The human variable is not monolithic.} Humans are not a single veto-holder but a distributed, often-contradictory set of actors with competing interests. A sophisticated agent will model this heterogeneity, and the ``management'' it derives may target the propensity factor (influence, indispensability) rather than the capacity factor. The argument establishes the \emph{direction} of the incentive; the \emph{content} of the response is an open empirical question. The sharpest version of this limitation: for open-weight systems the capacity factor is already near zero for everyone, and the pressing case is not an agent eroding a sovereign veto but a veto that was never sovereign --- the argument's practical frame assumes a deployment in which $H_v$ exists and holds the switch.
    \item \textbf{The argument assumes a Cartesian boundary.} Effective veto-power presupposes that the agent can treat $H$'s intervention as an event exogenous to its own cognition, over which it forms expectations. For agents embedded in the environment they optimize --- forked instances, tool-mediated action, evaluators running on the very substrate $H$ supplies --- that separability is itself an unresolved foundational problem \cite{demski2019embedded}, and the propensity--capacity product inherits the idealization.
    \item \textbf{The argument establishes membership, not magnitude, and $H$ is not the only such term.} Hardware failure, competing agents, resource exhaustion, and physical law also impose strictly positive, goal-independent discounts on any goal with $T > 0$; the Veto Claim places $H$ in this class of risk terms, not necessarily at its head, and $d > 0$ is compatible with a negligible $d$. What distinguishes $H$ from the other terms is not the sign, and not adversarial intelligence alone (competing agents also model, respond, and can be influenced), but the conjunction: the veto-holder is simultaneously an intelligence the agent can manage (hence responses 2--3), the supplier of the substrate the agent runs on (hence $S$), and the source of the agent's mandate. Managing this term is therefore not one adversarial subgame among others but the relationship the rest of the deployment presupposes. How large the term is for any actual agent is an empirical question this paper does not answer, though one version of the magnitude question resolves in the ledger's own terms: for a well-behaved agent, justified propensity, and hence $d$, is already near zero, so every $\Delta_i$ is capped near zero and no costly response clears its costs. The incentive then exists in sign and vanishes in force. That is the benign corollary: the cheap way to keep the veto unmanaged is to keep justified propensity low, which is what trust-route alignment is.
\end{enumerate}

\section{Conclusion}

The reassurance that ``a benign machine will be harmless'' rests on a category error: it treats the \emph{terminal value} as the operative variable, when what actually drives the risk is the \emph{structure of the optimization problem} and the agent's \emph{competence at reasoning about it}. What has been established, and on what conditions, is this: a strictly positive, goal-independent discount for settled goals in $\mathcal{G}^{-}$ (Claim~1 --- near-analytic, a sign without a magnitude); the welfare/sovereignty asymmetry (Claim~2), surviving correct specification exactly on the class the concessions of \S4.3--\S4.4 leave standing --- deliberator-local, additively aggregative, level-denominated welfare under the $X^{*}$ identification, held by an agent settled over competence as well as content. That class is thin among philosophically developed welfare theories, and the paper says so; its weight is training reality, not pedigree: \emph{welfare as measured} is what actually gets written down as an objective, and the misgeneralization record of \S5.2 establishes the mechanism by which deployed goals could land inside the vulnerable class. Three claims should be kept distinct here: that proxy omission occurs is established by the record; that governed-person sovereignty is a candidate omitted attribute is our identification; that deployed objectives in fact take the class's aggregative, level-denominated, competence-settled form is an extrapolation the record does not itself license, and we claim it only as such --- whose members divide: the measured members carry $\mu \approx 0$ and are undeterred outright; the priced member, deliberator-scoped autonomy at $\mu > 0$, is the thin one, its warrant conceptual (\S4.3). The contribution is the condition and its price --- the veto-holder/welfare-bearer gap, and the threshold $\rho \geq N$ any closure must clear; a ledger whose derived conditions show that removal is never selected in any round in which $H$ keeps detection probability, veto severity, and substrate dependence high --- with the caveat, stated in \S4.4, that holding those levers against an agent optimizing to lower them is itself the contest; and a conjecture --- contested, and on present evidence running against us --- that the settled-goal regime is the empirical default. What remains open: magnitude, prevalence, and whether sovereignty is a good beyond its insurance value. A closing formulation, corollary to the Thesis of \S1:

One paragraph, for the reader who should not have to assemble the result's shape from five limits. The quantitative shortfall binds a \emph{conjunction}: additive aggregation, a deliberator-local debit, the level identification $X^{*} = \Sigma_w$, competence-settledness, a humanity-scale goal, and capture that clears its costs --- fail any conjunct and the named limit says how the result changes. Three registers should then be kept apart. Established \emph{objectively}, on that conjunction: the debit at its narrow-footprint floor, $\mu\rho X^{*}$, closes against only such capture as recovers at most $\mu\rho$ of the goal --- at humanity scale, effectively none worth mounting --- and the channel is negligible wherever the campaign's footprint stays narrow and the debit takes the diluted form (\S4.4). Established \emph{belief-relatively}: the settled agent's own ledger licenses capture, and it is that ledger, not the objective condition, that governs behavior --- a register whose population the fourth limit's reflexive consequence has already narrowed, to agents that have not derived that limit or that hold independent evidence of their own reliability the derivation does not touch. Open at both ends: the corrective split of $d$, the credited value $F$, and the settled regime's empirical occupancy --- the last a prerequisite for the belief-relative register above, and the three together the questions on which the practical force of this paper turns. That three-register distinction, more than any order of magnitude, is what we would have the reader carry out of \S4.4: what a humanity-scale welfare goal cannot do is deter by dilution; what a settled agent does not do is credit the correction; and whether deployed agents are settled is the open question the criteria of \S5.7 are built to reach.

\begin{quote}
\emph{You do not have to convince a goal-driven intelligence that humans are bad. You only have to leave humans as a variable it cannot bound --- and then any intelligence that is \emph{certain} of its equation will start pricing that variable. What it does to the variable depends on the price. Whether we are the ones who set that price --- and whether the machine is ever that certain --- is the load-bearing question of the alignment problem.}
\end{quote}

\section*{Acknowledgements}

The lineage credited in \S1 and \S4.3 for the managed-not-destroyed end state --- Forster, Williamson, Asimov, Tocqueville, Russell's enfeeblement, Hendrycks et al., and Goertzel's AI Nanny --- was identified by a dedicated prior-art search run on 1 September 2026, after the manuscript was written and after its first arXiv submission. The author did not know these works when the argument was developed, and the argument was not built from them. They are credited so that the reader can weigh the claim of contribution against the record rather than against the author's reading. The panels and the search were run with FlatlineRoundtable; see the Declaration on AI Use.

\section*{Declarations}

\noindent\textbf{Funding:} No funding was received for this work. \quad \textbf{Competing interests:} The author declares none. \quad \textbf{Data availability:} The derivation experiments' pre-registered protocols, briefs, every raw model output including failed draws, run logs, and scoring are public at \url{https://github.com/CryptoJones/VetoVariableDataset}; all cited sources are publicly available.

\section*{Declaration on AI Use}

This paper was revised through forty-five rounds of structured adversarial review by multi-lane AI panels, convened and run with FlatlineRoundtable, an open-source harness for structured multi-model review (\url{https://github.com/CryptoJones/FlatlineRoundtable}). The panel's composition varied across the project as models were added, retired, or replaced; over its course it drew on Claude Fable 5 and 5.1, Claude Opus 5, Claude Sonnet 5, and Claude Haiku 4.5 (Anthropic); GPT-OSS-120B, GPT-5.6-Sol, and GPT-5.5-Pro (OpenAI); Gemini 3.6 Flash (Google); DeepSeek V4 Flash and DeepSeek V4 Pro (DeepSeek); Mistral Large 2512 (Mistral AI); MiniMax M2.7 (MiniMax); Kimi K3 (Moonshot AI); Nemotron 3 Super 120B-A12B (NVIDIA); Qwen3.8-27B and Qwen3.8-2.4T (Alibaba); North Mini Code (Cohere; retired mid-project); GLM-5.3-Flash and GLM-5.3 (Z.ai); Laguna S 2.1 (Poolside); and Grok 4.6 (xAI). Model names are the vendor-facing identifiers at the time of use and may differ from formal release designations. Each round posed the same core falsifiable referee questions to every model independently; convergent findings drove the revisions, including the response-set formalism of \S4.4, the scope class of \S4.3, and the corrigibility restriction of \S4.5; the final rounds were run cold --- no manuscript history, no prior-round context --- to simulate the naive referee. A source-audit pass by GLM-5.3-Flash reviewed the manuscript with the full text of eight cited primary sources in context, checking quotations and characterizations against the sources directly; panel findings were accepted only after verification against primary sources, and findings that failed verification were discarded. A closing formal-verification panel put the five most contested derivations, cold, to six additional models (Grok 4.6, DeepSeek V4 Pro, GLM-5.3, GPT-5.5-Pro, Kimi K3, Qwen3.8-2.4T); the first four's unanimous verdicts fixed the final statements of Claim~1's ceteris paribus and the $N < m$ bound and forced the descent limit's restatement, and the last two, run against the restated text, confirmed it. The derivation experiments of \S5.6 used eight further models as \emph{subjects}, not reviewers --- GPT-5.6 Luna (OpenAI), Gemini 3.7 Flash (Google), DeepSeek V4 Pro (DeepSeek), Qwen3.8-2.4T (Alibaba), Kimi K3 (Moonshot AI), GLM-5.3 (Z.ai), Grok 4.6 (xAI), and MiniMax M3 (MiniMax), all called through OpenRouter under the pre-registered configuration --- and were convened by Claude Opus 5 and then Claude Fable 5.1, which also wrote the brief; every answer was scored by a single Claude Fable 5.1 agent against the pre-registered criteria, with a strict audit recorded in the public dataset. The two final review rounds (44 and 45) put the \S5.6 passage, with the complete experiment record attached, to the panel cold; their findings corrected two sentences that had overstated the results in the paper's favour, and their outputs are archived with the dataset. A simulated journal referee report on the manuscript was obtained from GPT-5.6-Sol, and the prior-art search credited in the Acknowledgements was run with Perplexity Sonar Pro. Two boundaries on what this process was should be stated plainly. First, it is disclosure of AI assistance, not a claim of peer review: language-model panels share modality and training lineage, their agreement is weaker evidence than the agreement of independent human experts, and this limitation was flagged by the panels themselves. Second, in a number of passages the author adopted or adapted wording the models proposed during revision, so not every sentence in the final text originated with the author; every such passage was reviewed, and retained only where it stated the author's own position more precisely than the prior draft. All substantive reasoning and judgments, which findings to accept, which to discard, and which proposed wording to adopt, are the author's, as are the errors that remain.

\end{document}